\documentclass{article}

\PassOptionsToPackage{numbers, compress}{natbib}

 \usepackage[main, final]{neurips_2026}

\usepackage[utf8]{inputenc}
\usepackage[T1]{fontenc}
\usepackage{hyperref}
\usepackage{url}
\usepackage{booktabs}
\usepackage{amsfonts}
\usepackage{nicefrac}
\usepackage{microtype}
\usepackage{xcolor}

\usepackage{amsmath}
\usepackage{amssymb}
\usepackage{mathtools}
\usepackage{amsthm}
\usepackage{url}
\usepackage{pgfplots}
\pgfplotsset{compat=1.14}
\graphicspath{ {./images/} }
\usepackage{bbm}
\usepackage{comment}
\usepackage{xcolor}
\usepackage{bm}

\usepackage{algorithm}
\usepackage{algorithmic}

\theoremstyle{plain}
\newtheorem{theorem}{Theorem}[section]

\theoremstyle{definition}
\newtheorem{definition}[theorem]{Definition}

\theoremstyle{remark}
\newtheorem{remark}[theorem]{Remark}

\title{Near-Linear Time Generalized Sinkhorn Algorithms for Bounded Genus Graphs}

\author{%
  David S.~Hippocampus\thanks{Use footnote for providing further information
    about author (webpage, alternative address)---\emph{not} for acknowledging
    funding agencies.} \\
  Department of Computer Science\\
  Cranberry-Lemon University\\
  Pittsburgh, PA 15213 \\
  \texttt{hippo@cs.cranberry-lemon.edu} \\
}

\author{
Ananya Parashar, Derek Long, Dwaipayan Saha\\
Department of Industrial Engineering and Operations Research \\
Columbia University, New York, NY 10027\\
\texttt{\{ap4658,dl3538,ds4386\}@columbia.edu}
\And
Krzysztof Choromanski\\
Department of Industrial Engineering and Operations Research \\
Columbia University, New York, NY 10027\\
\texttt{\{ap4658,dl3538,ds4386\}@columbia.edu}
}

\author{\hspace{-1.mm}
Krzysztof Choromanski\textsuperscript{$1,2$ \thanks{equal contribution} $\,\,$\thanks{senior lead}$\,\,$}, Derek Long\textsuperscript{$1\,^{*}$}, Ananya Parashar\textsuperscript{$1\,^{*}$}, Dwaipayan Saha\textsuperscript{$1\,^{*}$}
\vspace{1.3mm}\\
\normalfont
\textsuperscript{$1$}Columbia University, 
\textsuperscript{$2$}Google DeepMind
\vspace{-1.3mm}
}

\begin{document}

\maketitle

\begin{abstract}
We present \textit{GenusSink}, a new class of approximate generalized Sinkhorn algorithms with shortest-path-distance costs for bounded genus (e.g. planar) graphs, providing near-linear time: (1) pre-processing, (2) iteration step, (3) final transport plan matrix querying and near-linear memory. Graphs handled by GenusSink include in particular planar graphs and bounded-genus meshes approximating 3D objects. GenusSink addresses total quadratic time complexity of its brute-force counterpart by leveraging separator-based decomposition of graphs, computational geometry techniques, and new results on fast matrix-vector multiplications with generalized distance matrices, using, in particular, Fourier analysis and low displacement rank theory. It is inspired by recent breakthroughs in graph theory on approximating bounded genus metrics with small treewidth metrics \citep{minor-free-paper}. The graph-centric approach enables us to target optimal transport problem with the corresponding distributions defined on the manifolds approximated by weighted graphs and with cost functions given by geodesic distances. We conduct rigorous theoretical analysis of GenusSink, provide practical implementations, leveraging newly introduced in this paper \textit{separation graph field integrators} (S-GFIs) data structures and present empirical verification. 
GenusSink provides orders of magnitude more accurate computations than other efficient Sinkhorn algorithms, while still guaranteeing significant computational improvements, as compared to the baseline. As a by-product of the developed methods, we show that GenusSink is \textbf{numerically equivalent} to the brute-force geodesic Sinkhorn algorithm on $n$-vertex graphs with treewidth $O(\log \log (n))$ (e.g. on trees).  
\end{abstract}

\section{Introduction \& Related Work}
\label{sec:intro}
Consider two spaces $X, Y$, with the corresponding measures $\mu$ and $\nu$ and a  function: $c: X \times Y \rightarrow [0, +\infty]$, quantifying the cost of transporting a ``unit of mass'' from locations in $X$ to those in $Y$. The celebrated \textit{Optimal Transport Problem} (OTP) \citep{pereira,montesuma2023recent,peyre2019computational,peyre2025otml,IG-OTP, so-otp, Figalli2010TheOP, villani, ambrosio} can be depicted as the task of ``transporting'' $\mu$ to $\nu$, by minimizing total cost. More formally, consider a measure $\pi \in \mathcal{P} (X \times Y)$, such that the projection of $\pi$ into $X$ and $Y$ results in $\mu$ and $\nu$ respectively, i.e.:
\begin{align}
\begin{split}
\pi(\mathcal{A} \times Y) = \mu(\mathcal{A}), \textit{    } \pi(X \times \mathcal{B}) = \nu(\mathcal{B}) \textit{    }\textrm{for all measurable   } \mathcal{A} \in X, \mathcal{B} \in Y.
\end{split}
\end{align}
We denote the set of such measures $\pi$ as $\Pi(\mu,\nu)$. The famous Kantorovich formulation \citep{kantorovich1942translocation} casts the OTP as the following optimization problem:
\begin{equation}
\label{eq:kantorovich}
\min_{\pi \in \Pi(\mu,\nu)} \int_{X \times Y} c(x,y)d\pi(x,y)    
\end{equation}
Optimal Transport Problem found applications in several fields of machine learning (ML), ranging from representation and metric learning \citep{frogner2015learning,metric-learning-2}, graph modeling \citep{peyre2016gromov,vayer2019optimal,xu2019gromov}, through computer vision \citep{kolkin2019style, rabin2011wasserstein} to generative modeling \citep{gulrajani2017improved,salimans2018improving, arjovsky2017wasserstein}, differentiable combinatorial computations \citep{diff-comb} and (more recently) even Transformer architectures \citep{espformer, kan}.

Without loss of generality, we can assume that $\mu$ and $\nu$ correspond to the probabilistic distributions. After the discretization, the optimization problem from Eq. \ref{eq:kantorovich} takes the following form: 
\begin{equation}
\label{eq:kantorovich_discrete}
\min_{\mathbf{P} \in \mathbb{R}^{n \times n}} \mathbf{P} \odot \mathbf{C} \textrm{  st.  } \mathbf{P}\mathbf{1}_{n} = \mathbf{a}, \textrm{  } \mathbf{P}^{\top}\mathbf{1}_{n}=\mathbf{b}, \textrm{  } \mathbf{P} \geq 0,   
\end{equation}
where $\mathbf{a} \in \mathbb{R}^{n}_{\geq 0},\mathbf{b} \in \mathbb{R}^{n}_{\geq 0}$ encode discrete probabilistic distributions ($\sum_{i=1}^{n}a_{i}=1$, $\sum_{j=1}^{n} b_{j}=1$).
In Eq. \ref{eq:kantorovich_discrete}, $\odot$ denotes Hadamard (element-wise) product, $\mathbf{C}=[c_{i,j}] \in \mathbb{R}^{n \times n}$ is called the \textit{cost matrix} and matrix $\mathbf{P}$, discretizing the joint distribution $\pi(X \times Y)$, is often referred to as a \textit{transport plan}. 

Matrix $\mathbf{P}^{*}$, encoding the optimal transport plan, already takes space $O(n^{2})$. However, in several applications, one only needs to efficiently query $\mathbf{P}$, e.g. compute efficiently $\mathbf{P}\mathbf{v}_{\mathcal{S}}$ for indicator vectors $\mathbf{v}_{\mathcal{S}}$, corresponding to subsets $S \subseteq \{1,...,n\}$ (providing mass amounts taken to a particular subset of target locations from initial locations). The brute-force, linear-program approach to solve the optimization problem from Eq. \ref{eq:kantorovich} (either finding $\mathbf{P}^{*}$ or efficient querying) requires cubic time complexity.

\textit{Sinkhorn (entropic) re-formulation} \citep{sinkhorn-1, sinkhorn-2} of the OTP relaxes it by adding entropy regularization term. This leads to the following optimization problem, where $\epsilon$ quantifies the importance of the regularizer:
\begin{equation}
\label{eq:entropic_form}
\min_{\mathbf{P} \in \mathbb{R}^{n \times n}} \mathbf{P} \odot \mathbf{C} + \epsilon \sum_{i,j}P_{i,j}(\log(P_{i,j})-1).  
\end{equation}
In the \textit{generalized $h$-Sinkhorn formulation}, $\log$ from Eq. \ref{eq:entropic_form} is replaced by $h$ for some $h:\mathbb{R} \rightarrow \mathbb{R}$.
Adding an extra term to the regular OTP formulation makes the problem convex and leads to this closed-form formula for $\mathbf{P}^{*}$, where $f=h^{-1}$ stands for $h$-inverse (for regular Sinkhorn, $f=\exp$):
\begin{equation}
\label{eq:k_matrix}
\mathbf{P}^{*} = \mathrm{diag}(\mathbf{u}) \cdot \mathbf{K}^{\epsilon}_{f} \cdot \mathrm{diag}(\mathbf{v}), \textrm{   } \mathbf{K}^{\epsilon}_{f} = f(-\frac{\mathbf{C}}{\epsilon}),    
\end{equation}
with vectors $\mathbf{u},\mathbf{v} \in \mathbb{R}^{n}$ obtained via the following iterative procedure: $\mathbf{v}^{0}=\mathbf{1}_{n}$, $\mathbf{u} \leftarrow \frac{\mathbf{a}}{\mathbf{K}^{\epsilon}_{f}\mathbf{v}}$,
$\mathbf{v} = \frac{\mathbf{b}}{(\mathbf{K}^{\epsilon}_{f})^{\top}\mathbf{u}}$. The exponentiation and division operations are conducted element-wise. Since in practice, the number of the steps of that iterative procedure (\textit{number of Sinkhorn iterations}) is constant, the overall time complexity of all Sinkhorn steps is quadratic, rather than cubic, in $n$. This algorithm still needs to materialize the cost matrix $\mathbf{C} \in \mathbb{R}^{n \times n}$, which takes quadratic space and at least quadratic time. Several computational improvements of the Sinkhorn algorithms were proposed \citep{sinkhorn-2, alaya2019screenkhorn, tang, Kacprzak, mengyu}. For instance, the \textit{Greenkhorn} algorithm \citep{sinkhorn-2}, improves the $O(n^{2})$ time complexity of the single iteration to $O(n)$, via efficient updates involving only one column/row of $\mathbf{K}$ per step. Those methods still require quadratic memory to store $\mathbf{C}$ and quadratic time to query $\mathbf{P}^{*}$ for accurate approximation. They also do not provide exact computations for low treewidth graph domains.

Consider the setting where $Y=X$. Without loss of generality, we assume that after the discretization, we have $n$ locations supporting both distributions. A natural class of cost functions for the OTPs in the non-Euclidean input spaces $X$ are shortest-path distances in the weighted undirected graphs approximating them. The above methods can be directly applied there, but at some point require the materialization of the so-called \textit{generalized distance matrix} $\mathbf{D}^{\epsilon}_{f} \in \mathbb{R}^{n \times n}$ defined as: $\mathbf{D}^{\epsilon}_{f} = [f_{\epsilon}(d_{i,j})]_{i,j=1,..,n}$, for $f_{\epsilon}:\mathbb{R} \rightarrow \mathbb{R}$ given as: $f_{\epsilon}(x) = f(-\frac{x}{\epsilon})$, and where: $d(i,j)$ stands for the shortest path between nodes: $i$ and $j$ in the given $n$-vertex graph $G_{X}$ that approximates $X$.

For general graphs $G_{X}$, it is hopeless to get more computationally-efficient methods, but one can ask whether we can do better under additional assumptions on $G_{X}$, present in several practical ML applications. In this paper, we provide a strong affirmative answer to this question.

We present \textit{GenusSink},  a new class of  approximate generalized Sinkhorn algorithms with shortest-path-distance costs for \textbf{bounded genus graphs} (e.g. planar graphs), providing near-linear time: (1) pre-processing, (2) iteration step, (3) final transport plan matrix querying and near-linear memory. Graphs handled by GenusSink include in particular \textbf{planar graphs} and \textbf{bounded genus} meshes approximating 3D objects. GenusSink addresses total quadratic time complexity of its brute-force counterpart by leveraging separator-based decomposition of graphs, computational geometry techniques, and new results on fast matrix-vector multiplications with generalized distance matrices, using in particular Fourier analysis and low displacement rank theory. It is inspired by recent breakthroughs in graph theory on approximating bounded genus metrics with small-treewidth metrics \citep{minor-free-paper}. The graph-centric approach enables us to target optimal transport problem with the corresponding distributions defined on the manifolds approximated by weighted graphs and with cost functions given by geodesic distances. We conduct rigorous theoretical analysis of GenusSink, provide practical implementations, leveraging \textit{separation graph field integrators} (S-GFIs) data structures, which are newly introduced in this paper, and present empirical verification. 
GenusSink provides orders of magnitude more accurate computations than other efficient Sinkhorn algorithms, while still guaranteeing significant computational improvements, as compared to the baseline.
As a by-product of the developed methods, we show that GenusSink is \textbf{numerically equivalent} to the brute-force geodesic Sinkhorn algorithm on the $n$-vertex graphs with treewidth $O(\log \log (n))$ (e.g. on trees). 

\begin{figure}
    \centering
    \includegraphics[width=0.9\linewidth]{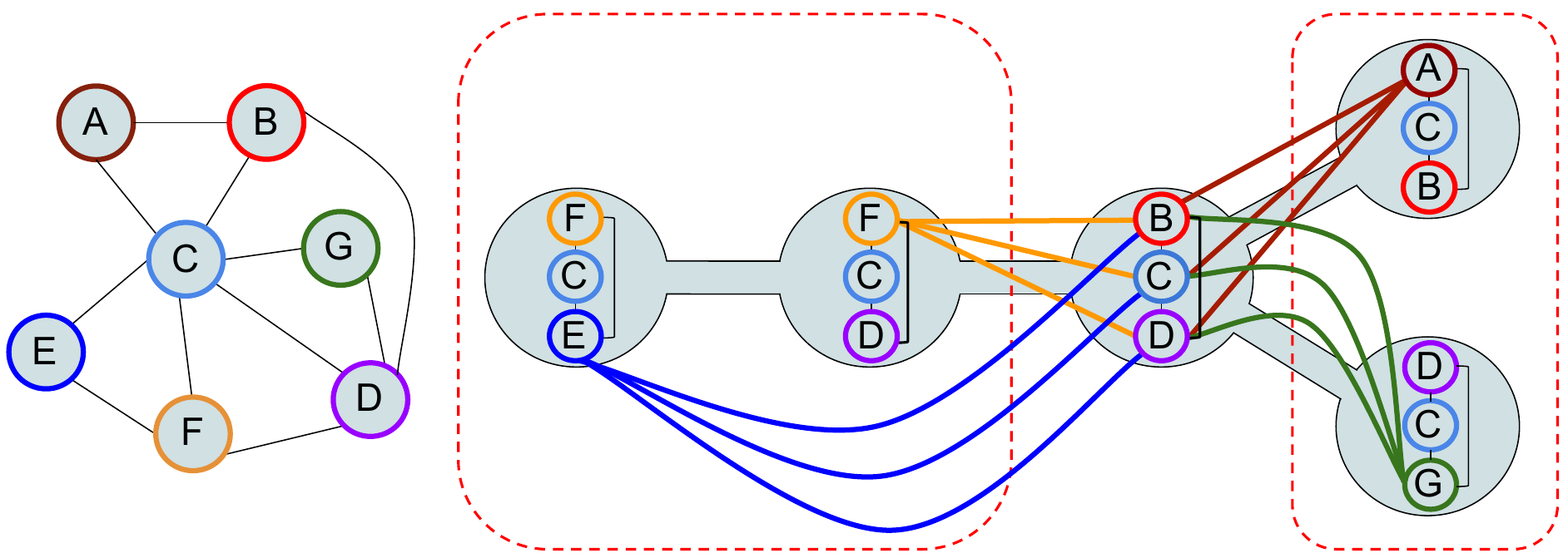}
    \caption{\small{\textbf{Left}: Seven-vertex planar graph. \textbf{Right:} A pictorial description of its \textit{treewidth} decomposition, with the treewidth parameter upper-bounding the sizes of the clusters of nodes above (that we will refer to as \textit{bags}, see: Sec. \ref{sec:method_treewidth}). Two graph pieces, obtained with the separator $\{B,C,D\}$, are highlighted via red dashed boxes. The treewidth decomposition provides a gateway for designing efficient algorithms  for various graph problems. In this paper, we leverage recent results in graph theory \citep{minor-free-paper} showing that the metrics of bounded genus graphs (in particular planar graphs, arising in several ML applications) can be accurately approximated by those of low treewidth graphs and the latter can be constructed in near-linear time. We apply this to propose a near-linear time Sinkhorn algorithm for distributions defined on the vertices of those graphs.}}
    \vspace{-4mm}
    \label{fig:treewidth}
\end{figure}

This paper is organized as follows (additional details are provided in the Appendix).

In Sec. \ref{sec:method_preliminaries}, we introduce basic graph theory concepts, used throughout this paper. Those include in particular the notion of the \textit{treewidth} (Sec. \ref{sec:method_treewidth}, see also: Fig. \ref{fig:treewidth}), a parameter describing how "close" a graph is to a tree (acyclic connected graph) and playing an important role in our algorithmic approach. In Sec. \ref{sec:method_sep_integrators} we present the separation graph field integration (S-GFI) data structure and show how it can be used to efficiently compute the actions of the generalized distance matrices for graphs under consideration. We then show how those translate to efficient Sinkhorn algorithms. In Sec. \ref{sec:method_treewidth_minor}, we finalize the GenusSink algorithm by lifting those techniques to bounded genus graphs, leveraging recent results on approximating metrics induced by bounded genus graphs with those of low treewidth graphs \citep{minor-free-paper}. In Sec. \ref{sec:method_practical}, we propose the practical implementation of GenusSink. Its empirical verification, including the comparison with \textbf{six} other efficient Sinkhorn methods and the brute-force, is conducted in Sec. \ref{sec:experiments}. We conclude in Sec. \ref{sec:conclusion}. Additional theoretical details regarding handling general functions $f$ are provided in the Appendix.

\section{Preliminaries}
\label{sec:method_preliminaries}

\begin{figure}
    \centering
    \includegraphics[width=0.9\linewidth]{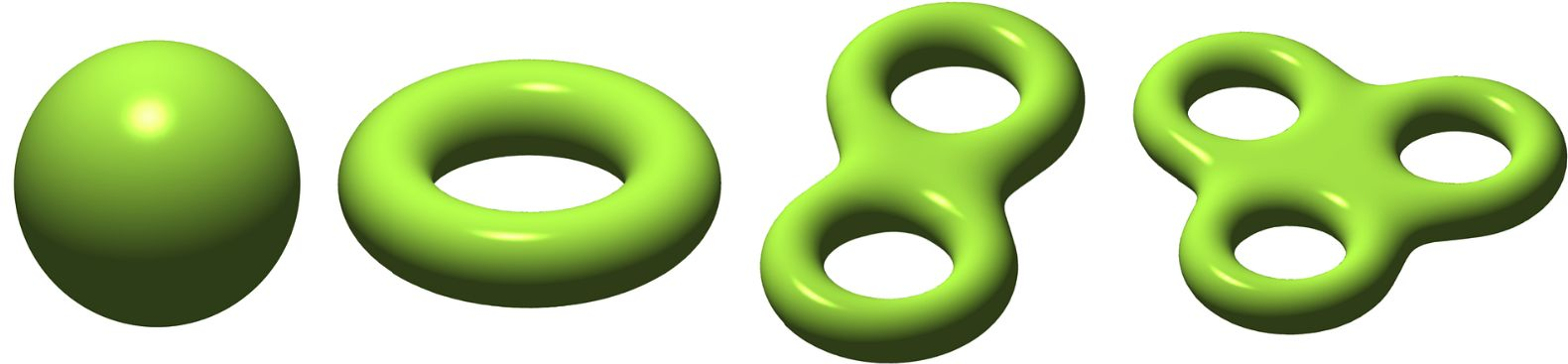}
    \caption{\small{Left to right: 2D-surfaces of 3D objects with genus values: $0$, $1$, $2$ and $3$. In this paper, we are especially interested in families of graphs with bounded genus that can be used in particular to provide discretized mesh representations of the above surfaces.}}
    \label{fig:genus}
\end{figure}
In this paper, we consider undirected weighted $n$-vertex graphs $G=(V,E,W)$ with the set of vertices $V=\{0,.,,,n-1\}$, the set of edges $E$ and the set of the corresponding positive weights $W$.
For a given subset $T \subseteq V$, we denote by $G[T]$ an \textit{induced subgraph} of $G$, obtained by taking vertices $T$ and the weighted edges between those vertices in $G$. We call a triplet $(A,S,B)$, for three pairwise disjoint subsets: $A,S,B \subseteq V$, satisfying: $A \cup S \cup B = V$, a \textit{separation configuration} if there are no edges between $A$ and $B$. We call $S$ a \textbf{separator} (vertices from $A$ and $B$ can be connected via paths of length $>1$ in $G$, but each such path needs to use a vertex from $S$). $S$ is called \textbf{balanced} if $|A|,|B| \geq cn$ for some fixed constant $c>0$, and where $|\cdot|$ stands for the set size.    

A graph is called \textbf{planar} if it can be drawn on the plane with non-intersecting edges. 
A surface of genus $g$ is a surface with $g$ "handles" (holes).
The \textbf{genus of the graph} $G$ is the smallest nonnegative integer $g$ such that $G$ can be drawn on an orientable surface of genus $g$ without edge crossing. We will be particularly interested in graphs of bounded genus (planar graphs provide a special sub-class here with zero-genus; see Fig. \ref{fig:genus}). 
\subsection{Treewidth \& Treewidth Decomposition}
\label{sec:method_treewidth}
The notion of the \textit{treewidth} provides a way to quantify how ``close'' the graph is to a tree. Since several problems that are challenging on general graphs can be efficiently solved for trees, this parameter effectively provides an upper bound on the "algorithmic tractability" of the graph (see also: Fig. \ref{fig:treewidth}).   

\begin{definition}[tree decomposition]
If $G=(V,E,W)$ is a weighted undirected graph, then its \textit{tree decomposition} is a pair $(T,\{X_{t}\}_{t \in V(T)})$, where $T$ is a tree with vertex-set $V(T)$ and sets $X_{t} \subseteq V$ are called \textit{bags} and satisfy the following properties:
\begin{itemize}
\item $\bigcup_{t \in V(T)} X_{t} = V$ (the bags cover all the vertices),
\item for each edge $(u,v) \in E$, there exists a bag $X_{t}$ such that: $\{u, v\} \subseteq X_{t}$,
\item for every vertex $v \in V$, the set $\{t \in V(T): v \in X_{t}\}$ is a sub-tree of $T$. 
\end{itemize}
\label{def: treewidth}
\end{definition}
The \textit{width} of a given tree decomposition $(T,\{X_{t}\}_{t \in V(T)})$ is defined as $\max_{t \in V(T)} |X_{t}|-1$, where \textbf{treewidth} $tw(G)$ of $G$ is the minimum width over all possible tree decompositions of $G$.
Directly from Def. \ref{def: treewidth}, we conclude that bags of the treewidth decomposition of $G$ act as separators in $G$. Not only are bags valid separators, but a bag that is a balanced separator can be found efficiently in time $O(n+|V(T)|)$, using \textit{centroid argument} for trees \citep{cygan}. Furthermore, the following holds \citep{diestel, Bodlaender}: 

\begin{theorem}
\label{theorem:treewodth}
For a given graph $G$ with $tw(G)=k$, there exists an algorithm of time complexity $2^{O(k)}n$ that outputs a tree decomposition $\mathcal{T}$ of width  $\leq 2k+1$. Furthermore, any tree decomposition of width $w$ can be converted into a tree decomposition of the same width and $O(w \cdot n)$ bags in linear time. Thus it can be assumed that $\mathcal{T}$ has $O((2k+1)n)$ bags.
\end{theorem}
Even though finding an optimal tree decomposition can be challenging, the above theorem shows that if we accept a width within a small multiplicative constant from optimal, this can be done efficiently. This assumes that the treewidth is very small, as compared to $n$. That will be indeed the case in our analysis, where we will work with graphs of treewidth $O(\log \log n)$, whenever we apply those techniques. We can also assume that such a decomposition does not have too many bags. This will enable us to efficiently find those bags of the decomposition that are balanced separators. 
\section{GenusSink Algorithm}
\label{sec:method}
\subsection{Separation Graph Field Integrators (S-GFI) for Sinkhorn}
\label{sec:method_sep_integrators}
The separation graph field integrator (S-GFI) data structure provides a way to efficiently compute matrix-vector multiplications with generalized distance matrices $\mathbf{D}^{\epsilon}_{f}=[f_{\epsilon}(d_{i,j})]_{i,j=1,,,.n}$ for several classes of functions $f_{\epsilon}:\mathbb{R} \rightarrow \mathbb{R}$, and graphs with small treewidth, where $d_{i,j}$ stands for the shortest path distance between nodes $i,j$ in a given graph $G$. This plays a critical role in designing GenusSink.

\subsubsection{An Overview}
The name: \textit{graph field integrator} (GFI) corresponds to the fact that the action $\mathbf{D}^{\epsilon}_{f}\mathbf{x}$ of a matrix $\mathbf{D}^{\epsilon}_{f}$ on a given vector $\mathbf{x} \in \mathbb{R}^{n}$ can be thought of as a process of integrating a scalar field on the vertices of $G$ and given by $\mathbf{x}$. The integration coefficients depend on the distances between the nodes in the graph (shortest path distance) metric. On a high-level, S-GFI is a tree with nodes corresponding to balanced separators (to apply divide-and-conquer strategy) and additional metadata supporting fast integration. 

In the first phase, S-GFI is built in the \textbf{one-time} process per graph. That construction critically relies on the ability to efficiently find small balanced separators. It also leverages Dijkstra's shortest-path algorithm \citep{cormen} to construct meta-data useful for the efficient calculations of those contributions to the overall integration that arise from sub-matrices $\mathbf{D}_{f_{\epsilon}}^{A,B} = [f_{\epsilon}(d_{i,j})]_{i \in A, j \in B} \in \mathbb{R}^{|A| \times |B|}$, for sets $A, B$ corresponding to the separation $(A,S,B)$ (nodes of the S-GFI tree correspond to separators $S$).

In the second phase, when S-GFI is ready, the integration is conducted for any $\mathbf{x} \in \mathbb{R}^{n}$. The key component of that procedure is a method computing efficiently actions of the matrices $\mathbf{D}_{f_{\epsilon}}^{X,Y} = [f_{\epsilon}(d_{i,j})]_{i \in X, j \in Y} \in \mathbb{R}^{|X| \times |Y|}$
for $X=A,Y=B$ and $X=B,Y=A$. Here we apply computational geometry methods. For generalized distance matrices, we also leverage Fourier analysis and low displacement rank theory (see: Appendix, Sec. \ref{sec:appendix_cross}), but in the regular Sinkhorn setting with $f_{\epsilon}:x\rightarrow \exp(-\frac{x}{\epsilon})$, this is not needed. See Fig. \ref{fig:sgfi} (left) for the visual scheme of S-GFI.

\subsubsection{Deeper dive into S-GFI data structure}
\label{sec:sgfi-deepdive}
For a given graph $G$ and subsets $X, Y \subseteq V(G)$ with $X \subseteq Y$, we denote by $G^{X}[Y]$ a subgraph of $G$ induced by the subset $Y$, but with the edges between the nodes of $X$ modified as follows. For any two different $i,j \in X$ a weight of an edge between $i$ and $j$ is assigned as the shortest path distance between $i$ and $j$ in $G$ (thus if necessary, new edges are added; if $i$ and $j$ are not connected via any path in $G$, no edge is added). S-GFI is a binary tree with the following node-attributes (and separation denoted as $(A,S,B)$):
\begin{itemize}
\item \textbf{\textrm{sep\_dist\_matrix}}: matrix of shape $|S |\times |S|$ of shortest path distances between the nodes of the small balanced separator $S$ of the graph represented by a sub-tree rooted at that node,    
\item \textbf{\textrm{left\_distances}}: a matrix of shape $|A| \times |S|$, containing shortest path distances between vertices of $A$ and vertices of $S$,
\item \textbf{\textrm{right\_distances}}: a matrix of shape $|B| \times |S|$, containing shortest path distances between vertices of $B$ and vertices of $S$,
\item \textbf{\textrm{left/sep/right\_ids}}: the ids of the vertices from $A/S/B$,
\item \textbf{\textrm{explicit\_graph}}: an explicit graph that the sub-tree rooted at that node corresponds to (this is $\textrm{None}$, unless a node is a leaf).
\item \textbf{\textrm{left/right\_sgfi}}: left/right child of the node, corresponding to graph $G^{S}[A \cup S]$ / $G^{S}[B \cup S]$.
\end{itemize}
S-GFI supports efficient multiplication $\mathbf{D}^{\epsilon}_{f}\mathbf{x}$ via the \textbf{$\textrm{integrate}$} function, given in Algorithm 1. The critical component of the algorithm is a procedure: \textbf{\textrm{cross\_compute}} that given two matrices of distances: $\mathbf{AD} = \textbf{\textrm{left\_distances}} \in \mathbb{R}^{|A| \times |S|}$ and $\mathbf{BD} = \textbf{\textrm{right\_distances}} \in \mathbb{R}^{|B| \times |S|}$, as well as two vectors: $\mathbf{u} \in \mathbb{R}^{|A|}$ and $\mathbf{v} \in \mathbb{R}^{|B|}$, computes: $\mathbf{M}^{A,B}_{f_{\epsilon}}\mathbf{v}$ and $(\mathbf{M}^{A,B}_{f_{\epsilon}})^{\top}\mathbf{u}$ in near-linear time, where $\mathbf{M}_{f_{\epsilon}}^{A, B}$ is defined as follows: 
\begin{equation}
\mathbf{M}^{A,B}_{f_{\epsilon}} = \left[f_{\epsilon}\left(\min_{k=1,...,|S|} (\mathbf{AD}[i][k]+\mathbf{BD}[k][j])\right)\right]_{i=1,...,|A|}^{j=1,...,|B|} \in \mathbb{R}^{|A| \times |B|}   
\end{equation}
Since all paths between vertices of $A$ and $B$ use a vertex from $S$ (by the definition of the separator), we conclude that $\mathbf{M}^{A,B}_{f_{\epsilon}} = \mathbf{D}^{A,B}_{f_{\epsilon}}$. Thus vectors $\mathbf{L},\mathbf{R}$ from l.5 of Algorithm 1 
encode contributions to $\mathbf{y}=\mathbf{D}^{\epsilon}_{f}\mathbf{x}$ from its two sub-matrices: $\mathbf{D}^{A,B}_{f_{\epsilon}}$ and $\mathbf{D}^{B,A}_{f_{\epsilon}} = (\mathbf{D}^{A,B}_{f_{\epsilon}})^{\top}$.

\begin{figure}
    \centering
    \includegraphics[width=0.99\linewidth]{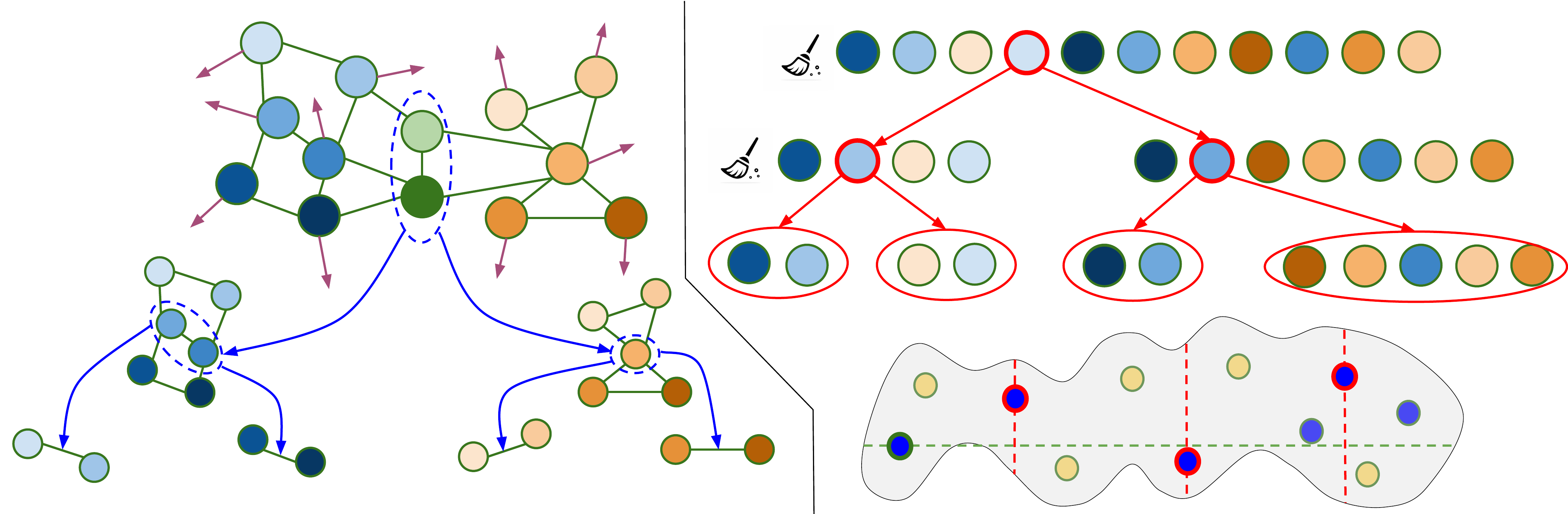}
    \caption{\small{\textbf{Left:} The pictorial representation of the S-GFI data structure. In that example, the corresponding tree has a root with two children and four grandchildren. \textbf{Right:} The visualization of the \textbf{\textrm{cross\_compute}} function with the medians depicted in red/green and their corresponding splits highlighted as dashed lines.}}
    \label{fig:sgfi}
\end{figure}

\begin{algorithm}[t]
\caption{\textbf{\textcolor{violet}{Integrate method:}} Calculate efficiently $\mathbf{y}=\mathbf{D}^{\epsilon}_{f}\mathbf{x}$ for a given $\mathbf{x} \in \mathbb{R}^{n}$, using S-GFI}\label{alg:integrate}
\textbf{Input:} separation graph field integrator $\textrm{SGFI}$, input vector $\mathbf{x} \in \mathbb{R}^{n}$ \\
\textbf{Output:} Exact or approximate $\mathbf{y}=\mathbf{D}^{\epsilon}_{f}\mathbf{x}$ \\
\begin{algorithmic}[1]
\STATE initialize: $y=\textbf{0} \in \mathbb{R}^{n}$
\IF{$\textrm{SGFI}.\textrm{explicit\_graph}$ is None}
    \STATE $\textrm{AD} \leftarrow \textrm{SGFI}.\textrm{left\_distances}$, 
            $\textrm{AID} \leftarrow \textrm{SGFI}.\textrm{left\_ids}$
    \STATE $\textrm{BD}  \leftarrow \textrm{SGFI}.\textrm{right\_distances}$,
            $\textrm{BID} \leftarrow \textrm{SGFI}.\textrm{right\_ids}$
    \STATE $\mathbf{L}, \mathbf{R} \leftarrow \textrm{SGFI}.\textrm{cross\_compute}(\textrm{AD}, \textrm{BD}, \mathbf{x}[\textrm{AID}]$, $\mathbf{x}[\textrm{BID}]$)
    \STATE $\mathbf{y}[\textrm{AID}] \textrm{+=} \mathbf{L}$, $\mathbf{y}[\textrm{BID}] \textrm{+=} \mathbf{R}$, $\textrm{SID} \leftarrow \textrm{SGFI}.\textrm{sep\_ids}$
    \STATE $\textrm{ASID} \leftarrow \textrm{sort}(\textrm{concat}(\textrm{AID}, \textrm{SID}))$, $\textrm{BSID} \leftarrow \textrm{sort}(\textrm{concat}(\textrm{BID}, \textrm{SID}))$
    \STATE $\textrm{ASY} = \textrm{integrate}(\textrm{SGFI}.\textrm{left\_sgfi},\mathbf{x}[\textrm{ASID}])$, $\textrm{BSY} = \textrm{integrate}(\textrm{SGFI}.\textrm{right\_sgfi},\mathbf{x}[\textrm{BSID}])$
    \STATE $\mathbf{y}[\textrm{ASID}] \textrm{+=} \textrm{ASY}$, $\mathbf{y}[\textrm{BSID}] \textrm{+=} \textrm{BSY}$, $\textrm{FS} \leftarrow f_{\epsilon}(\textrm{SGFI}.\textrm{sep\_dist\_matrix})$
    \STATE $\textrm{SY} = \textrm{FS} \cdot \mathbf{x}[\textrm{SID}]$, $\mathbf{y}[\textrm{SID}] \textrm{-=} \textrm{SY}$
\ELSE
    \STATE compute $\mathbf{D}^{\epsilon}_{f}\mathbf{x}$ brute-force using $\textrm{SGFI}.\textrm{explicit\_graph}$
\ENDIF
\STATE return $\mathbf{y}$
\end{algorithmic}
\end{algorithm}
Assuming that this computation can be done efficiently, the remaining part of Alg. 1 is a relatively straightforward application of the divide-and-conquer strategy. If the node of SGFI contains an explicit graph it corresponds to, the computation is done brute-force (l.12). Otherwise the contributions to $\mathbf{y}$ from $\mathbf{D}^{A,B}_{f}$ and $\mathbf{D}^{B,A}_{f}$ are added in l.6, as discussed before. This is followed, by adding contributions from $\mathbf{D}^{A \cup S, A \cup S}_{f}$ and $\mathbf{D}^{B \cup S, B \cup S}_{f}$ via the recursive invocation of the procedure (l.9). Finally, in l.10 we subtract a contribution from $\mathbf{D}^{S,S}_{f}$ (computed brute-force), since it was counted twice in previous calculations.
The correctness of the algorithm follows from the fact that in the graph corresponding to the left and right child of the given SGFI node, edge-weights between vertices of $S$ are equal to the lengths of the shortest paths between them in $G$, if they exist (thus the lengths of the shortest paths between nodes $i, j \in A$ or $i, j \in B$ can be retrieved from the left or right child only, even if those paths intersect with both $A$ and $B$).
We present an efficient implementation of \textbf{\textrm{cross\_compute}} next.
\paragraph{Procedure $\textrm{cross\_compute}$:} Let the rows of $\mathbf{AD} \in \mathbb{R}^{|A| \times |S|}$ and $\mathbf{BD} \in \mathbb{R}^{|B| \times |S|}$ be: $\mathbf{a}_{1},...,\mathbf{a}_{|A|} \in \mathbb{R}^{|S|}$ and $\mathbf{b}_{1},...,\mathbf{b}_{|B|} \in \mathbb{R}^{|S|}$ respectively. We will show how to compute efficiently $\mathbf{M}_{f_{\epsilon}}^{A,B}\mathbf{v}$. The computation of $\mathbf{M}_{f_{\epsilon}}^{B,A}\mathbf{u}$ can be conducted analogously. 
Note that the following holds:
\begin{equation}
(\mathbf{M}^{A,B}_{f_{\epsilon}}\mathbf{v})[i] = \sum_{k=1}^{|S|} 
\sum_{j=1}^{|B|}f_{\epsilon}(\mathbf{a}_{i}[k]+\mathbf{b}_{j}[k])
\mathbf{I}[\mathbf{w}_{i}^{k} \prec \mathbf{z}_{j}^{k}]\mathbf{v}_{j},
\end{equation}
where $\prec$ denotes element-wise $<$-inequality and $\mathbf{w}^{k}_{i},\mathbf{z}^{k}_{j}$ are defined as follows:
\begin{align}
\begin{split}
\mathbf{w}^{k}_{i} = (\mathbf{a}_{i}[k]-\mathbf{a}_{i}[0],...,\mathbf{a}_{i}[k]-\mathbf{a}_{i}[k-1], \mathbf{a}_{i}[k]-\mathbf{a}_{i}[k+1],...,\mathbf{a}_{i}[k]-\mathbf{a}_{i}[|S|]) \in \mathbb{R}^{|S|-1} \\   
\mathbf{z}^{k}_{j} = (\mathbf{b}_{j}[0]-\mathbf{b}_{j}[k],...,\mathbf{b}_{j}[k-1]-\mathbf{b}_{j}[k], \mathbf{b}_{j}[k+1]-\mathbf{b}_{j}[k],...,\mathbf{b}_{j}[|S|]-\mathbf{b}_{j}[k]) \in \mathbb{R}^{|S|-1}
\end{split}    
\end{align}
We will show how to efficiently compute the expressions $\xi_{i} = \sum_{j=1}^{|B|}f_{\epsilon}(\mathbf{a}_{i}[k]+\mathbf{b}_{j}[k])
\mathbf{I}[\mathbf{w}_{i}^{k} \preceq \mathbf{z}_{j}^{k}]\mathbf{v}_{j}$ for a fixed $k$ and all $i$. We then repeat the process $|S|$ times to consider all $k$.
We start with the case $|S|=1$. If $f=\exp$ (regular Sinkhorn objective) then:
\begin{equation}
(\xi_{1},...,\xi_{|A|})^{\top} = \left(\exp(-\frac{\mathbf{a}_{1}[k]}{\epsilon}),...,\exp(-\frac{\mathbf{a}_{|A|}[k]}{\epsilon})\right)^{\top}
\left(\exp(-\frac{\mathbf{b}_{1}[k]}{\epsilon}),...,\exp(-\frac{\mathbf{b}_{|B|}[k]}{\epsilon})\right)\mathbf{v}
\end{equation}
and thus all $\xi_{i}$ can be clearly computed in time $O(|A|+|B|)$
(with the one time pre-computation of the dot-product $(\exp(-\frac{\mathbf{b}_{1}[k]}{\epsilon}),...,\exp(-\frac{\mathbf{b}_{|B|}[k]}{\epsilon}))\mathbf{v}$. 
Interestingly, near-linear (approximate or exact) algorithms exist for a much broader class of functions $f$. We provide detailed corresponding analysis in the Appendix, applying several additional techniques: Fourier analysis, random features and the theory of low displacement rank matrices. That enables us to extend the results presented in the main body of the paper to a generalized $h$-Sinkhorn setting.
Now let us assume that $|S|>1$. Then we sort all vectors: $\mathbf{w}^{k}_{1},...,\mathbf{w}^{k}_{|A|},\mathbf{z}^{k}_{1},...,\mathbf{z}^{k}_{|B|}$ in the ascending order of their first coordinates (see: Fig. \ref{fig:sgfi} (right)). We denote the corresponding sorted sequence as: $(\mathbf{t}_{1},...,\mathbf{t}_{|A|+|B|})$. We have:
$\mathbf{t}_{1}[0] \leq ... \leq \mathbf{t}_{|A|+|B|}[0]$.
We compute the median of the first coordinates of the $\mathbf{w}$-vectors from $(\mathbf{t}_{1},...,\mathbf{t}_{|A|+|B|})$. Let us denote by $\mathcal{L}^{A}$ the set of indices of those $\mathbf{w}$-vectors with the first coordinate upper-bounded by this median. Denote by $\mathcal{R}^{A}$ the set of indices of those $\mathbf{w}$-vectors with the first coordinate larger than this median. We define $\mathcal{L}^{B}$ and $\mathcal{R}^{B}$ analogously (using $\mathbf{z}$-vectors, but the median already computed for $\mathbf{w}$-vectors).
The task of computing all $\xi_{i}$ for $i=1,..., |A|$ reduces to this of computing all $\xi_{i}$ with $i \in \mathcal{L}^{A}$ and all $\xi_{i}$ with $i \in \mathcal{R}^{A}$.
We get the following for $i \in \mathcal{L}^{A}$:
\begin{equation}
\label{eq:cross-1}
\xi_{i} = \sum_{j \in \mathcal{L}^{B}}f(\mathbf{a}_{i}[k]+\mathbf{b}_{j}[k])\mathbf{I}[\mathbf{w}^{k}_{i} \prec \mathbf{z}^{k}_{j}]\mathbf{v}_{j} + \sum_{j \in \mathcal{R}^{B}}f(\mathbf{a}_{i}[k]+\mathbf{b}_{j}[k])\mathbf{I}[\mathbf{w}^{k}_{i}[1:] \prec \mathbf{z}^{k}_{j}[1:]]\mathbf{v}_{j}    
\end{equation}
Furthermore, for $i \in \mathcal{R}^{A}$, we obtain:
\begin{equation}
\label{eq:cross-2}
\xi_{i} = \sum_{j \in \mathcal{R}^{B}}f(\mathbf{a}_{i}[k]+\mathbf{b}_{j}[k])\mathbf{I}[\mathbf{w}^{k}_{i} \prec \mathbf{z}^{k}_{j}]\mathbf{v}_{j}  
\end{equation}
We can then continue the calculations recursively.
The key observation is that each of the sums from Eq. \ref{eq:cross-1}, \ref{eq:cross-2} needs to be computed for only half of the number of entries $i$ that the original sums needed to be computed for. Furthermore, second sum in Eq. \ref{eq:cross-1} uses sub-vectors $\mathbf{w}$ and $\mathbf{z}$ obtained by skipping their first dimensions. 
By solving standard recursive equation for the time complexity of the \textbf{$\textrm{cross\_compute}$} procedure, we conclude that it is: $O(n \cdot \textrm{poly}(|S|) \cdot \log^{\textrm{poly(|S|)}}(n))$. Using this fact and solving another standard recursive time complexity equation for Algorithm 1, we conclude that:
\begin{remark}
The overall time complexity of the procedure \textbf{$\textrm{integrate}$} from Algorithm 1 is $O(n \cdot \textrm{poly}(|S|) \cdot \log^{\textrm{poly(|S|)}}(n))$ for a polynomial function $\textrm{poly}$, separator-size $|S|$ and number of vertices $n$.
\end{remark}
It remains to build the S-GFI data structure.
Note that given a small balanced separator $S$, all other steps for the construction of the particular node of the S-GFI tree can be conducted straightforwardly with the classical Dijkstra's shortest path algorithm that runs in time $O(n+m)\log(n)$, where $m=O(n \cdot tw(G))$ is the number of edges of the input bounded genus graph (we leverage well-known upper bound on $m$, using $tw(G)$). Thus if calculating $S$ can be done in $O(n \cdot \textrm{poly}(|S|) \cdot \log^{\textrm{poly(|S|)}}(n))$ time, the overall time complexity of the construction is also: $O(n \cdot tw(G) \cdot \textrm{poly}(|S|) \cdot \log^{\textrm{poly(|S|)}}(n))$. We can also leverage the tree decomposition algorithm, discussed in Sec. \ref{sec:method_treewidth}. For the given input graph $G$ with treewidth $tw(G)=k$, we can first construct its tree decomposition $\mathcal{T}$, as in Theorem \ref{theorem:treewodth} in time $O(2^{O(k)})n$. We can then find a bag corresponding to a small balanced separator of size at most $2k+1$ in time $O(nk)$, as discussed in Sec. \ref{sec:method_treewidth}. Note that importantly, we do not need to re-calculate $\mathcal{T}$ as we proceed with the calculations of the children of the given S-GFI node, but simply use the corresponding subtrees of $\mathcal{T}$. Indeed, adding extra edges between the vertices of the bags (we do it for the balanced separator), \textbf{does not destroy} tree decomposition structure. We conclude that:
\begin{remark}
For graphs $G$ with $tw(G)=O(\log \log (n))$, building S-GFI can be conducted in near-linear time $O(n \cdot \textrm{poly}(\log \log (n)) \cdot \log^{\textrm{poly}(\log \log (n))}(n))$ (in particular $o(n^{1+\epsilon})$ for any $\epsilon>0$). The space complexity is also near-linear in $n$.  
\end{remark}
\subsubsection{From distance matrices to efficient Sinkhorn}
Given S-GFI that can be efficiently built and applied to multiply with generalized distance matrices, the path to the efficient Sinkhorn-algorithm with the following features: (1) efficient pre-processing, (2) improved iteration step and (3) improved querying of the final transport plan matrix (given a fixed number of Sinkhorn iteration steps), as well as: (4) efficient memory is straightforward. One only needs to build the S-GFI instance to support efficient multiplication with matrices $\mathbf{K}^{\epsilon}_{f}$ from Eq. \ref{eq:k_matrix}, that are exactly the generalized distance matrices we considered here. There is however a hidden assumption here that  $tw(G)=O(\log \log (n))$. We will address it next to finalize the algorithm. 
\subsection{From Small Treewidth Graphs to Bounded Genus Graphs}
\label{sec:method_treewidth_minor}
Bounded genus $n$-vertex graphs do not necessarily have treewidth of order $O(\log \log (n))$. For example, planar graphs have treewidth $O(\sqrt{n})$, but not necessarily smaller. However shortest-path metrics on bounded genus graphs can be accurately approximated by the $O(\log \log (n))$-treewidth proxies and furthermore, those proxies can be efficiently computed. The following holds:
\begin{theorem}[\citep{minor-free-paper}]
\label{theorem:main}
Given an $n$-vertex graph $G=(V,E,W)$ of diameter $D$ from the family of bounded genus graphs and a parameter $\epsilon \in (0,1)$, there is a deterministic embedding $f:V(G) \rightarrow H$ into a graph $H$ of treewidth $O(\epsilon^{-1}(\log \log (n))^{2})$ and additive distortion $+\epsilon D$. Furthermore, $f$ can be deterministically constructed in $O(n \frac{\log^{3}(n)}{\epsilon^{2}})$ time.
\end{theorem}
The original statement given in \citep{minor-free-paper} is for planar graphs, but the Authors explicitly show in the proof how the statement can be extended to bounded genus graphs. 

With Theorem \ref{theorem:main}, we can finalize GenusSink algorithm. For a given graph $G$ from the bounded genus family, we first construct its embedding into $H$, as given in Theorem \ref{theorem:main}. We then run our version of the Sinkhorn algorithm, applying S-GFI data structure on $H$. 
\subsection{Practical implementations}
\label{sec:method_practical}
We observed that in practical implementations, we do not even need to construct an embedding, as in Theorem \ref{theorem:main}. Instead we can run well-known linear time algorithms for constructing $O(\sqrt{n})$-size balanced separators
for graphs $G$ from the bounded genus family under consideration. We can then sub-sample a $\log \log (n)$ random points from each separator and distribute the remaining points of the separator randomly to both children of the given node of the S-GFI structure. As we show in Sec. \ref{sec:experiments}, that version is still very accurate, in particular orders of magnitude more accurate than other efficient Sinkhorn methods used on a regular basis. Finally, we do not need to construct all the levels of the S-GFI tree. The depth $h$ of the tree is effectively a hyperparameter, and we can terminate at any time, simply by storing explicit graphs in the corresponding nodes (via \textbf{\textrm{explicit\_graph}} fields). Hyperparameter $h$ provides a convenient tradeoff between asymptotic computational superiority of the methods presented in this paper and highly optimized brute-force matrix-vector multiplications (for graphs small enough regular Sinkhorn algorithm leveraging standard matrix-vector multiplication suffices). In practice, we applied small values of $h$.

\section{Experiments}
\label{sec:experiments}
\subsection{Surfaces and Meshes}
\begin{figure}[h]
    \centering
    \includegraphics[width=0.9\linewidth]{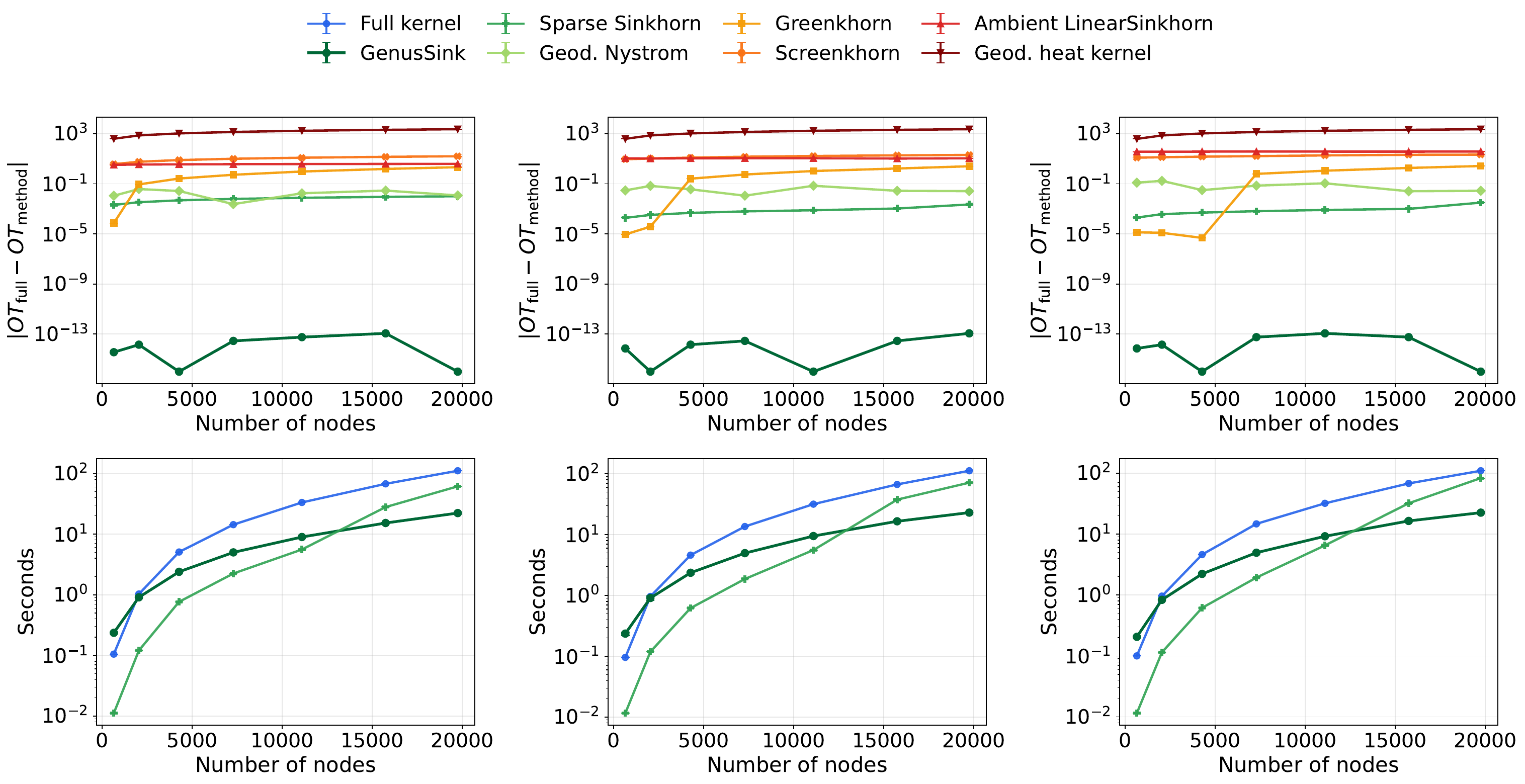}
    \caption{\small{Synthetic pseudo-genus surface experiments for pseudo-genus orders 1, 2, and 3 (left to right). \textbf{Top:} absolute Sinkhorn-cost error relative to the dense geodesic full-kernel baseline. \textbf{Bottom:} total runtime for the Brute Force (full kernel) and the two most accurate efficient methods from previous studies: GenusSink and Sparse Sinkhorn. GenusSink stays numerically exact while providing significant computational gains.}}
    \label{fig:pseudo-genus-accuracy-total-time-results}
\end{figure}

We evaluate GenusSink on a family of boundaryless sphere-topology ``pseudo-genus'' surface graphs that remain planar (see Fig.~\ref{fig:pseudo-genus-meshes-3d}) and compare GenusSink against a comprehensive set of baseline approximation methods \cite{alaya2019screenkhorn,sinkhorn-2,schmitzer2019stabilized,linear-time-sinkhorn-positive-features,huguet2023geodesic}—Appendix~\ref{app:details-simple} provides further details on these methods and the experiment set-up. Across all pseudo-genus families, GenusSink matches the geodesic full-kernel baseline to numerical precision while becoming significantly faster at larger scales (see Fig.~\ref{fig:pseudo-genus-accuracy-total-time-results}). In contrast, most approximate baselines exhibit a clear accuracy gap.

Extending to more challenging settings, we evaluate GenusSink on real meshes from the \textbf{Thingi10K} dataset \cite{Zhou2016Thingi10K} with up to \textbf{13k} vertices. We convert mesh edges to weighted graph edges using Euclidean lengths, and generate smooth source and target measures as geodesic Gaussian mixtures around deterministic PCA-based mesh anchors. 
Again, accuracy compared with other approximation methods shows that GenusSink remains essentially exact while avoiding full dense kernel materialization on these real mesh graphs (see Fig. \ref{fig:thingi10k-results}). Further experiment details can be found in Appendix~\ref{app:details-thingi10k}.

\begin{figure}[h]
    \centering
    \raisebox{0.3\height}{%
        \includegraphics[width=0.15\linewidth]{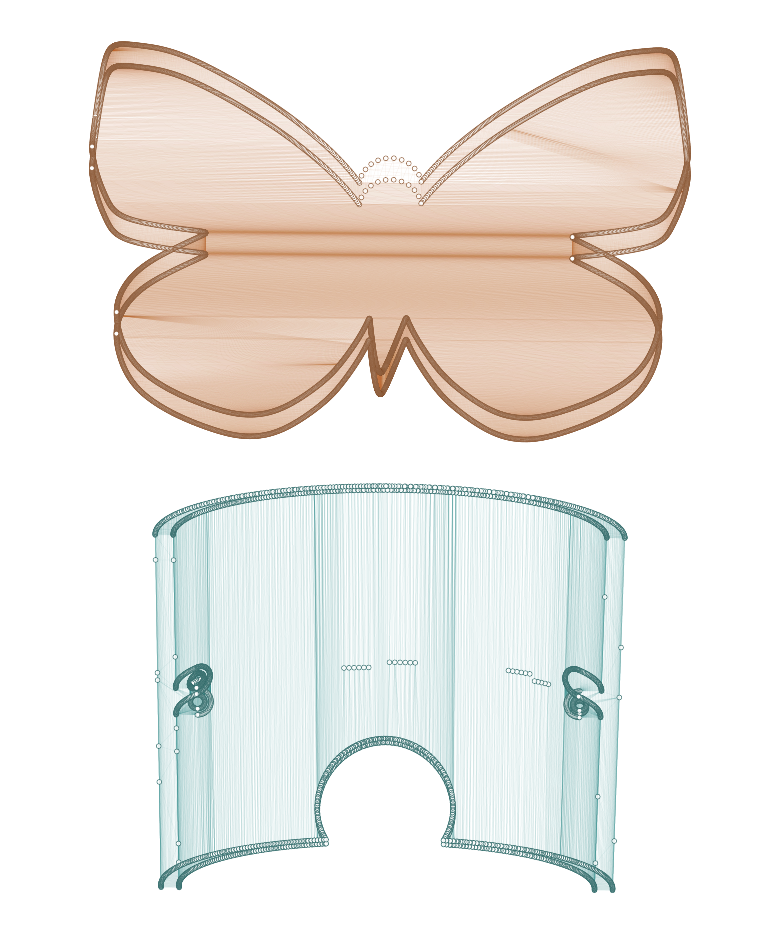}%
    }\hfill
    \includegraphics[width=0.41\linewidth]{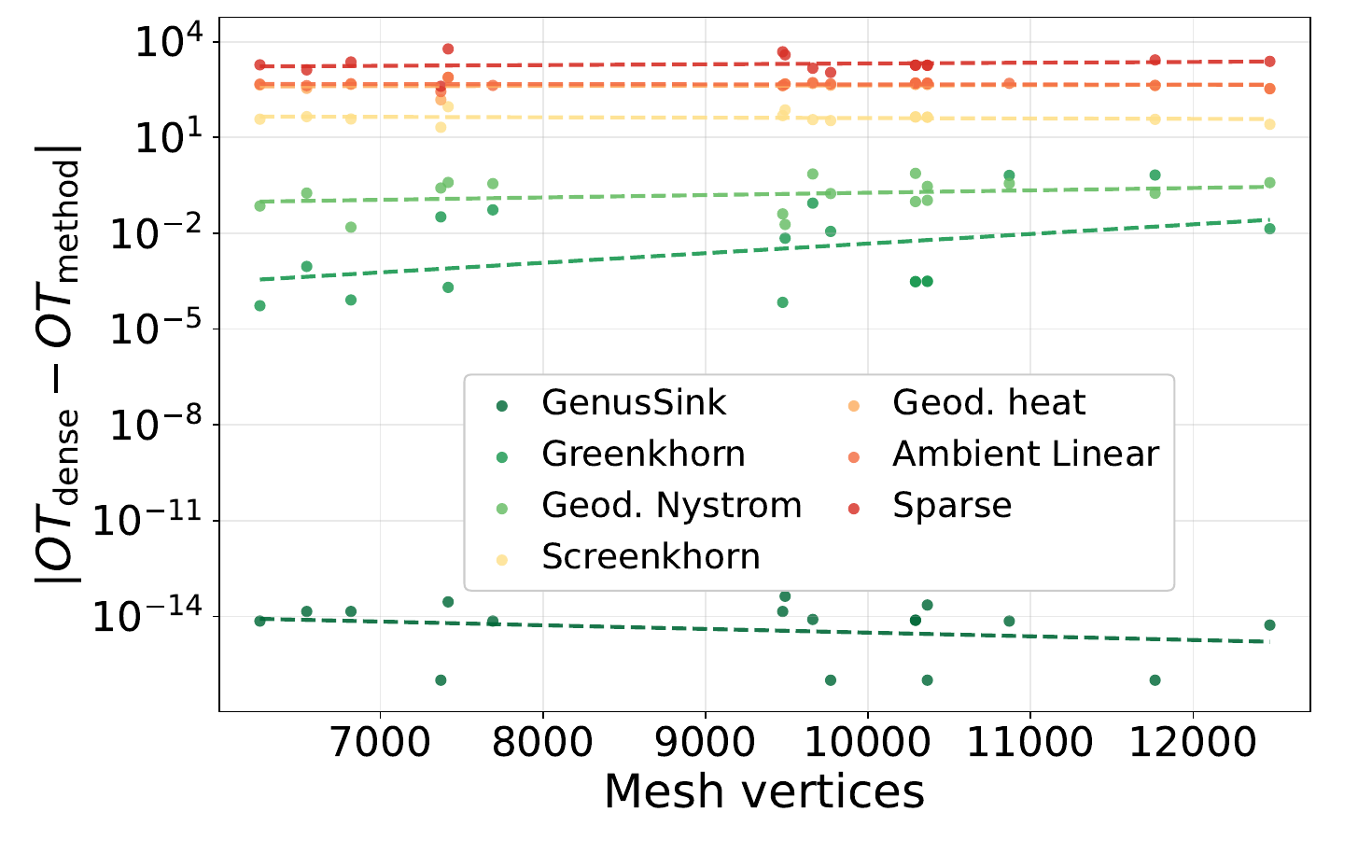}
    \includegraphics[width=0.41\linewidth]{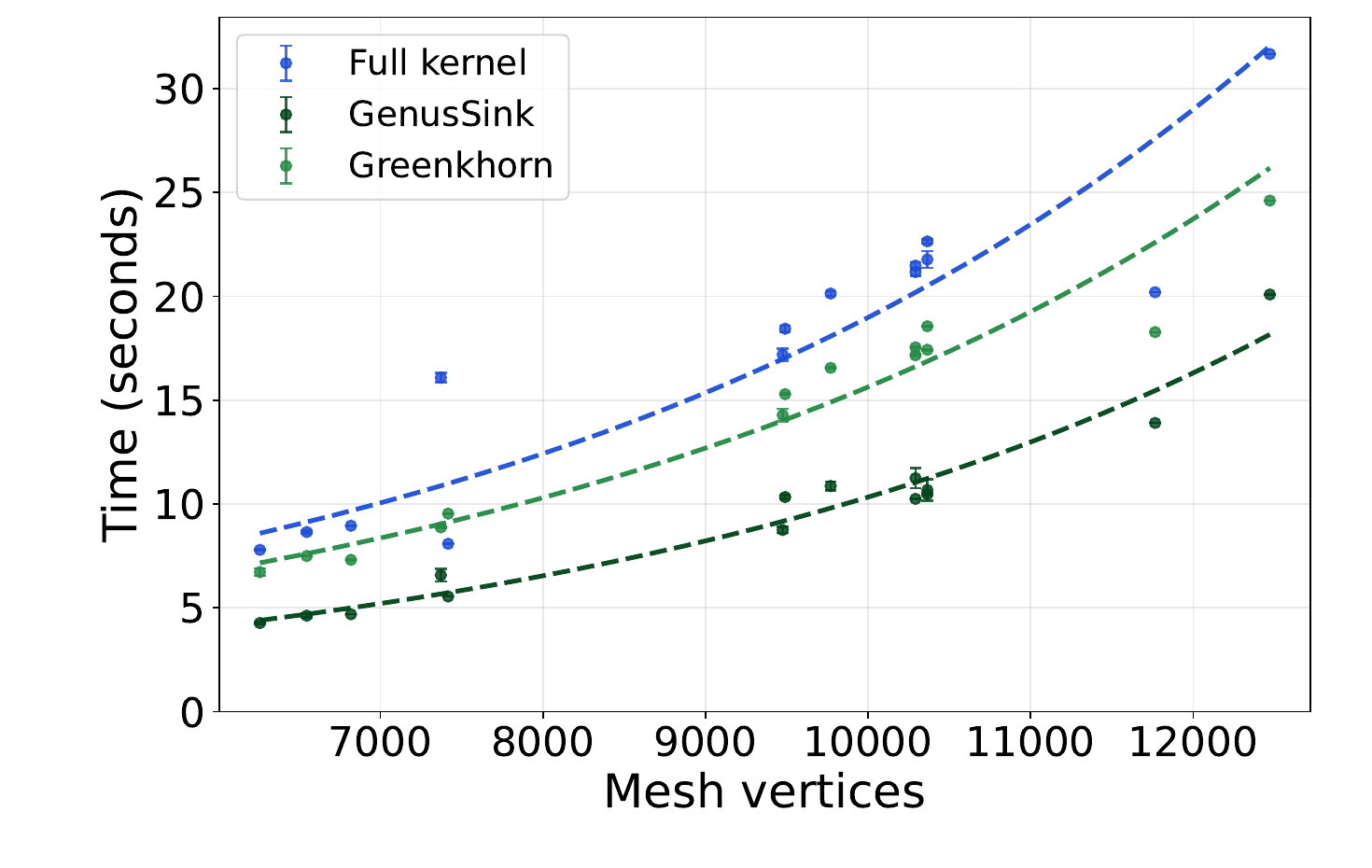}
    \caption{\small{\textbf{Left:} examples of Thingi10K meshes.  \textbf{Middle:} absolute OT-cost error relative to dense geodesic Sinkhorn for several efficient Sinkhorn methods. As before, GenusSink stays numerically exact while providing significant computational gains. \textbf{Right:} runtime scaling for dense geodesic kernel construction versus two most accurate efficient methods: GenusSink and Greenkhorn (see: left), with error bars over repeated runs.}}
    \label{fig:thingi10k-results}
\end{figure}

\subsection{Enhancing Ambulance Deployment in New York City}
Emergency Medical Services (EMS) is a high-stakes application where every minute can materially affect patient outcomes. New York City operates one of the largest EMS systems in the world, serving more than 1M incidents annually. We model EMS deployment as the problem of matching ambulance supply to emergency-call demand on a Bronx road graph with 33,363 nodes, where nodes represent possible ambulance locations and the demand distribution is induced by historical incident patterns from \href{https://data.cityofnewyork.us/Public-Safety/EMS-Incident-Dispatch-Data/76xm-jjuj/about_data}{publicly available dispatch data}. The goal is to choose ambulance locations that reduce mean response and upper-tail response time, two operationally important EMS metrics.

\begin{figure}[h]
    \centering
    \includegraphics[width=0.2\linewidth]{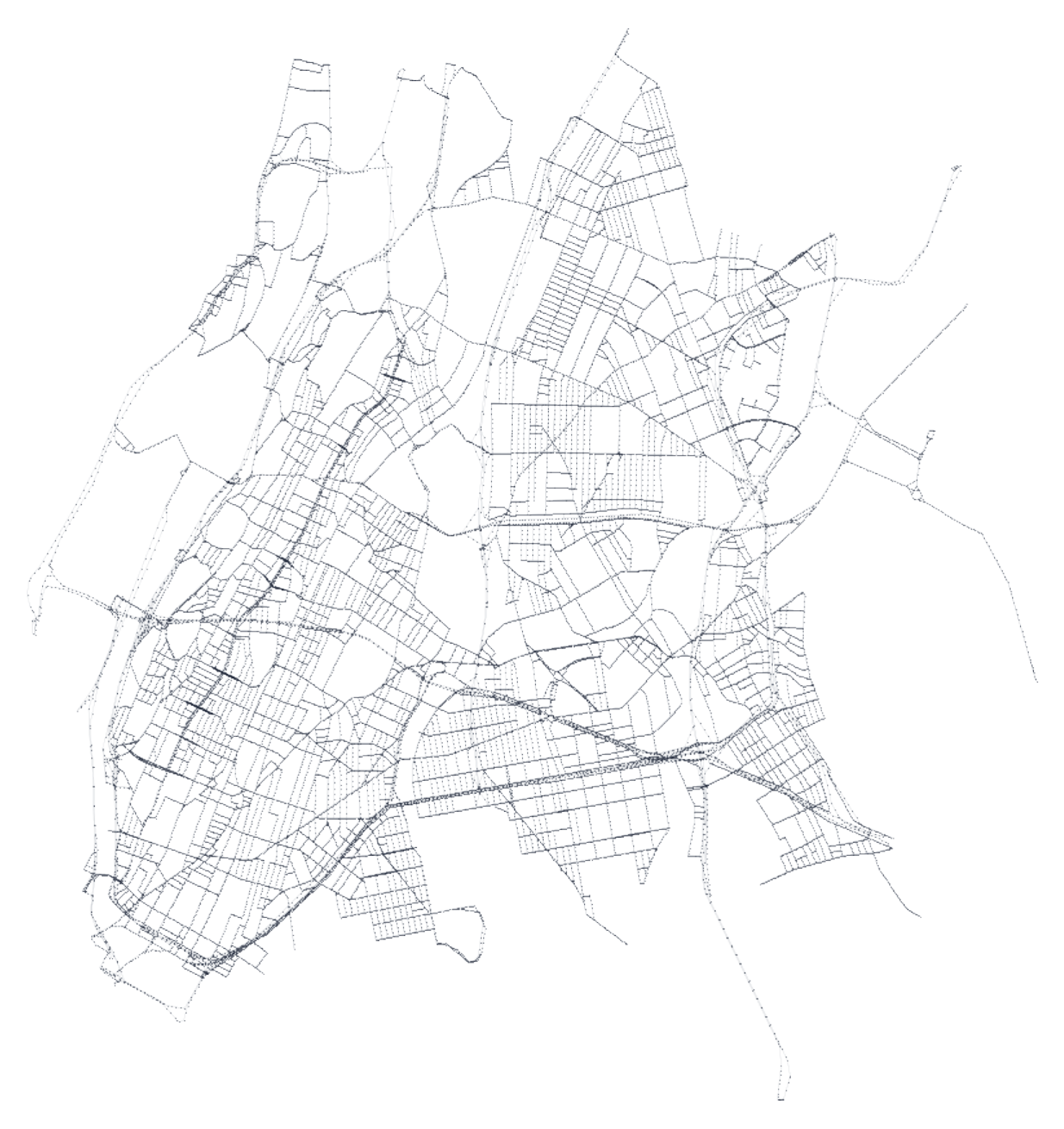}
    \includegraphics[width=0.39\linewidth]{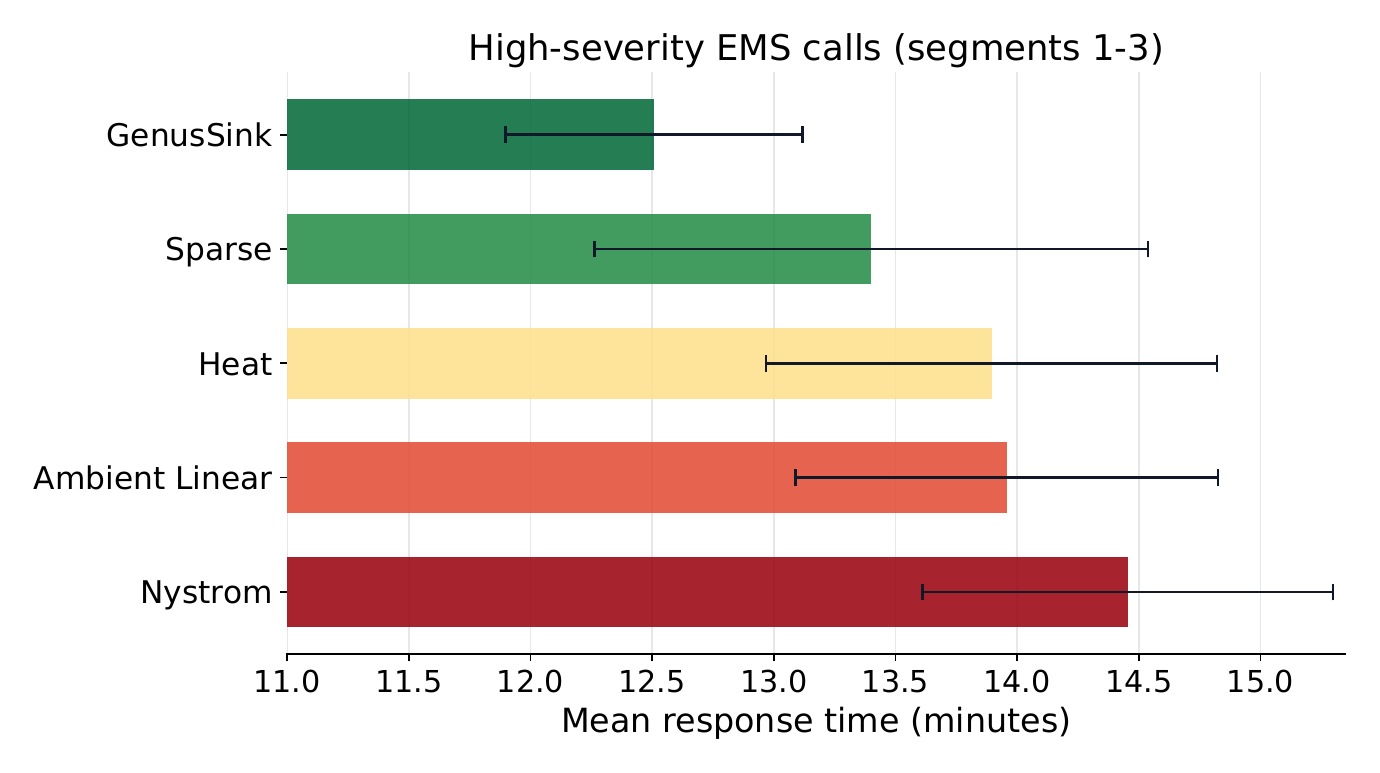}
    \includegraphics[width=0.39\linewidth]{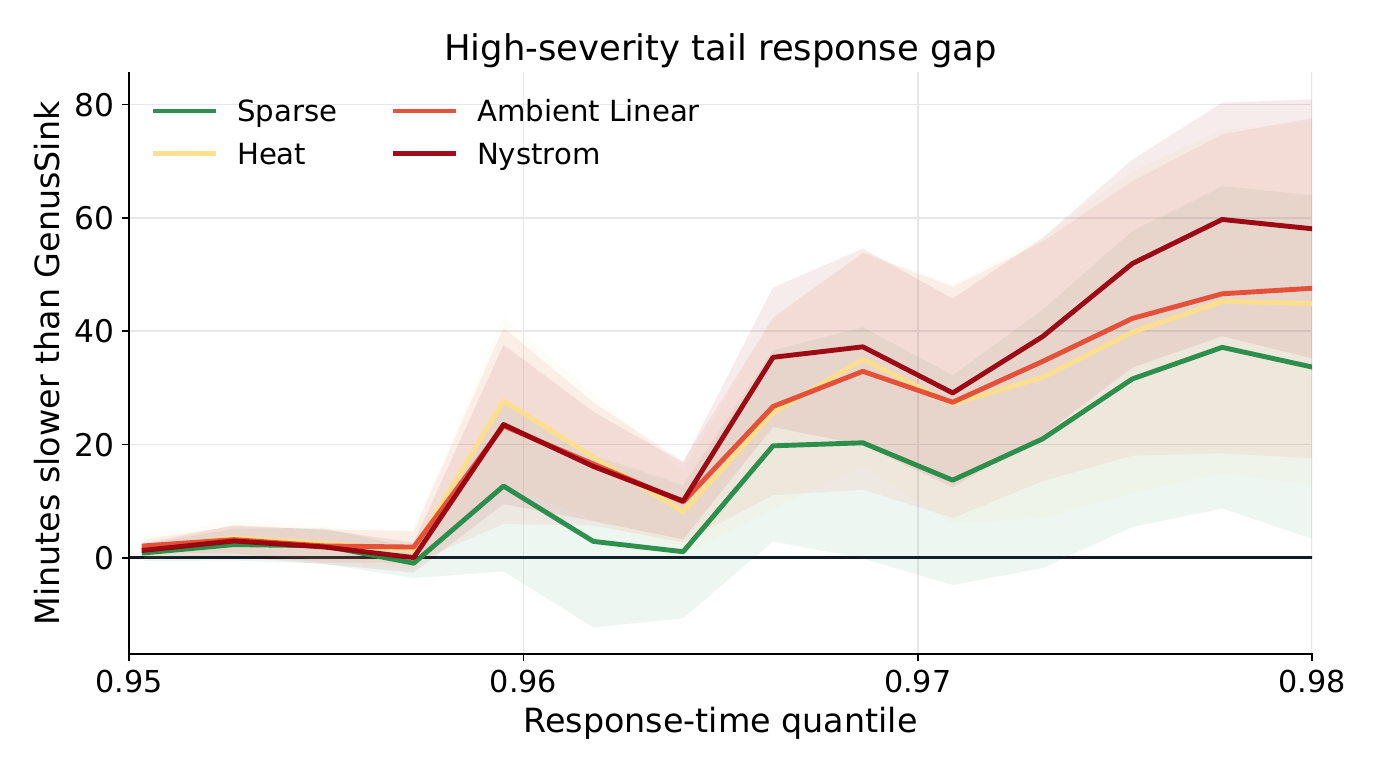}
    \caption{\small{Bronx road graph (left), mean response times across different Sinkhorn methods (middle), and tail response time gaps relative to GenusSink at quantiles from 0.95 onwards (right).}}
    \label{fig:ems-downstream-application-results}
\end{figure}

We evaluate this task using a well-established EMS simulator called JEMSS \cite{RIDLER20221101} and add a new configuration for the Bronx, a borough of particular operational concern for EMS decision-makers. In our experiment, given the empirical demand distribution, we deploy 60 ambulances and compare GenusSink against other approximation baselines. Note that computing and storing the full graph kernel on this network would require all-pairs road distances over tens of thousands of nodes, which is intractable for frequent ambulance redeployment and motivates the need for fast, accurate Sinkhorn approximations. Hence, we did not include baselines that require materializing the full geodesic kernel. We see that GenusSink consistently produces better deployments: for high-severity calls (i.e., segments 1--3), it achieves the lowest mean response time, 12.5 minutes, compared with 13.4 to 14.5 minutes for the approximation baselines (see Fig.~\ref{fig:ems-downstream-application-results}). The advantage is especially pronounced in the upper tail, where the baseline methods become increasingly slower than GenusSink as the response-time quantile increases beyond the 95th percentile.

\section{Conclusion}
\label{sec:conclusion}
We proposed in this paper \textit{GenusSink}, a new class of efficient Sinkhorn algorithms operating on graph metric spaces for bounded genus graphs and characterized by near-linear time (1) pre-processing, (2) iteration step, (3) final transport plan matrix querying and (4) near-linear memory. GenusSink leverages several techniques from graph theory and computational geometry. We provided detailed theoretical analysis of the algorithm and complemented it with thorough empirical evaluation.

\bibliography{refs}
\bibliographystyle{plainnat}

\appendix

\section{Efficient computation of \textrm{cross\_compute} for $|S|=1$ and general $f$}
\label{sec:appendix_cross}

The problem of computing efficiently \textbf{\textrm{cross\_compute}} is equivalent to the problem of the efficient (near-linear) multiplication with matrices $\mathbf{M}=[f(x_{i}+y_{j})]_{i=1,...,|A|}^{j=1,...,|B|}$ for given sequences of scalars: $(x_{1},...,x_{|A|})$ and $(y_{1},...,y_{|B|})$. In this section, we will define efficient (approximate or exact) algorithms to achieve this task. We start with the observation that the so-called \textit{cordial functions}, defined in \citep{ftfi}, by definition, support near-linear multiplication:

\begin{definition}[cordial functions; \citep{ftfi}]
A function $f:\mathbb{R} \rightarrow \mathbb{R}$ is $d$-cordial (or: cordial if $d$ is not specified) if there exists $d \in \mathbb{N}$ such that matrix-vector multiplication with a matrix 
$\mathbf{M}=\left[f(x_{i}+y_{j})\right]_{i=1,...,a}^{j=1,...,b}$ can be \textbf{exactly} conducted in time $O((a+b)\log^{d}(a+b))$ for any two sequences of scalars $(x_{1},...,x_{a})$ and $(y_{1},...,y_{b})$.
\end{definition}

As explained in \citep{ftfi}, several general classes of functions are cordial:
\begin{itemize}
\item Rational functions $f$ are $(2+\epsilon)$-cordial for any $\epsilon>0$. Efficient multiplication can be conducted with the use of Fast Fourier Transform (FFT).
\item Bounded-degree polynomial functions are $0$-cordial. Thus for polynomial functions (special classes of rational functions) the above bound can be improved.
\item Exponential functions are $0$-cordial. This is literally the result that we presented in the main body of the paper, since it corresponds to the regular Sinkhorn algorithm with the entropy-based regularizer.
\item Functions of the form: $f(x) = \frac{\exp(\lambda x)}{x+c}$ for hyperparameters $\lambda, c$ are $3$-cordial. This follows from the fact that in this case the corresponding matrix is the so-called \textit{Cauchy-type low displacement rank matrix} and the theory of low-displacement rank matrices \citep{ldr} (or to be more specific, \textit{Cauchy Transform}) can be applied to produce efficient multiplication algorithm of time complexity $O((a+b)\log^{3}(a+b))$. 
\end{itemize}

We can conclude the following:
\begin{remark}
For all the classes of functions $f$ defined above, \textit{GenusSink} for the generalized Sinkhorn problem with the corresponding function $f$ and leveraging the above methods remains asymptotic computational profile (space and time complexity wise) of its regular counterpart (for $f=\exp$).
\end{remark}





\subsection{Fourier analysis and random features for \textrm{cross\_compute}}
Assume that we can re-write function $f$ as:
\begin{equation}
f(\mathbf{x}+\mathbf{y}) \approx \eta_{1}^{\top}(\mathbf{x})\eta_{2}(\mathbf{y})    
\end{equation}
for some randomized $\eta_{1},\eta_{2}:\mathbb{R}^{d} \rightarrow \mathbb{R}^{m}$. Then it is easy to see that we can compute \textrm{cross\_compute} for the case $|S|=1$ in time $O(m(|A|+|B|))$, using matrix-associativity property. Thus for $m=\textrm{polylog}(|A|+|B|)$, the computations will be done efficiently.  
We will show now  a systematic method to construct $\eta_{1},\eta_{2}$. Note that we have:
\begin{equation}
f(\mathbf{z})=\int_{\mathbb{R}^{d}} \exp(-2\pi k \omega^{\top}\mathbf{z})\tau(\omega)d\omega,    
\end{equation}
where $\tau$ is the inverse Fourier Transform of $f$:
\begin{equation}
\tau(\omega) = \int_{\mathbb{R}^{d}} \exp(2 \pi k \mathbf{x}^{\top}\omega)f(\mathbf{x}) d\mathbf{x} 
\end{equation}
($k^{2}=-1$). Thus, taking: $\mathbf{z}=\mathbf{x}+\mathbf{y}$, we can rewrite:
\begin{equation}
f(\mathbf{x}+\mathbf{y})=C \cdot \mathbb{E}_{p(\omega)}
[\exp(2 \pi k \omega^{\top} \mathbf{x})
\exp(2 \pi k \omega^{\top} \mathbf{y})],    
\end{equation}
where $p(\omega)$ is the probabilistic distribution with density proportional to $\tau(\omega)$ and $C=\int_{\mathbb{R}^{d}} \tau(\omega) d\omega$ (here we assume that the latter integral is well-defined!).
Thus, if we can efficiently (potentially approximately) sample from $p(\omega)$, then we can leverage the following:
\begin{equation}
f(\mathbf{x}+\mathbf{y}) \overset{\mathbb{E}}{=} \eta^{\top}_{1}(\mathbf{x})(\eta_{2}(\mathbf{y})),    
\end{equation}
where for $\omega_{1},...,\omega_{m} \sim p(\omega)$ and for some $r \in \mathbb{N}$ (number of random features) the following holds:
\begin{align}
\begin{split}
\eta_{1}(\mathbf{v}) = \eta_{2}(\mathbf{v}) = \sqrt{\frac{C}{r}}
\left(\exp(2 \pi k \omega_{1}^{\top}\mathbf{v}),...,\exp(2 \pi k \omega_{r}^{\top}\mathbf{v})\right).
\end{split}
\end{align}
Thus the only thing we required from $f$ is to have a well defined Inverse Fourier Transform $\tau$ and also in principle for $\tau$ to satisfy: $\tau \geq 0$. But the last assumption is actually not needed since we can always partition $\tau$ into its positive and negative part and conduct calculations separately for each of them, only to combine them at the end.

We can conclude the following:
\begin{remark}
If $f$ has the well-defined Inverse Fourier Transform, \textit{GenusSink} for the generalized Sinkhorn problem with the corresponding function $f$ and leveraging the above method remains asymptotic computational profile (space and time complexity wise) of its regular counterpart (for $f=\exp$), as long as $m=\textrm{polylog}(|A|+|B|)$. Note that here \textbf{\textit{cross\_compute}} is approximated rather than computed exactly (as it was the case for all scenarios considered before).
\end{remark}

\section{Numerical Experiments}
\label{sec:num_experiments}
All experiments were run on an Apple M3 machine with 8 CPU cores, a 10-core integrated GPU, and 16 GB of unified memory, running macOS 26.3.1.

\subsection{Details on the pseudo-genus family}
\label{app:details-simple}

\begin{figure}[h]
    \centering
    \includegraphics[width=0.49\linewidth]{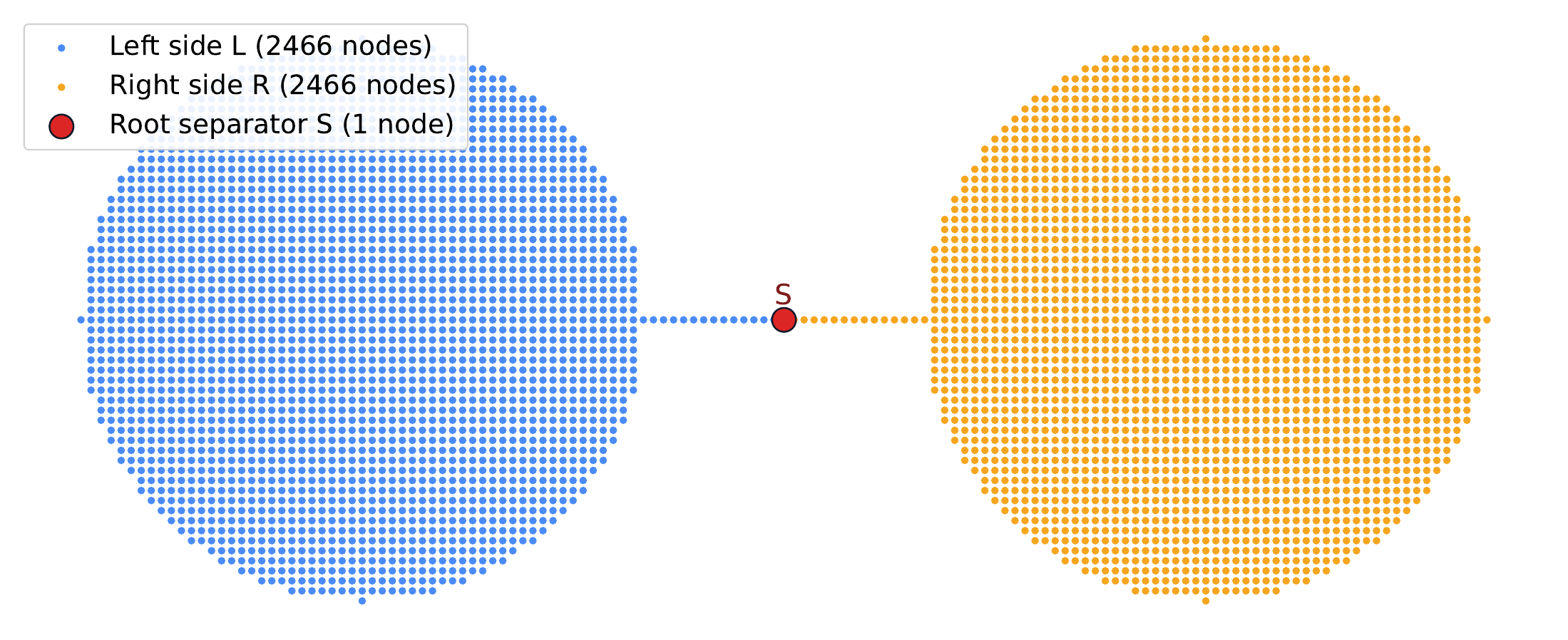}
    \includegraphics[width=0.49\linewidth]{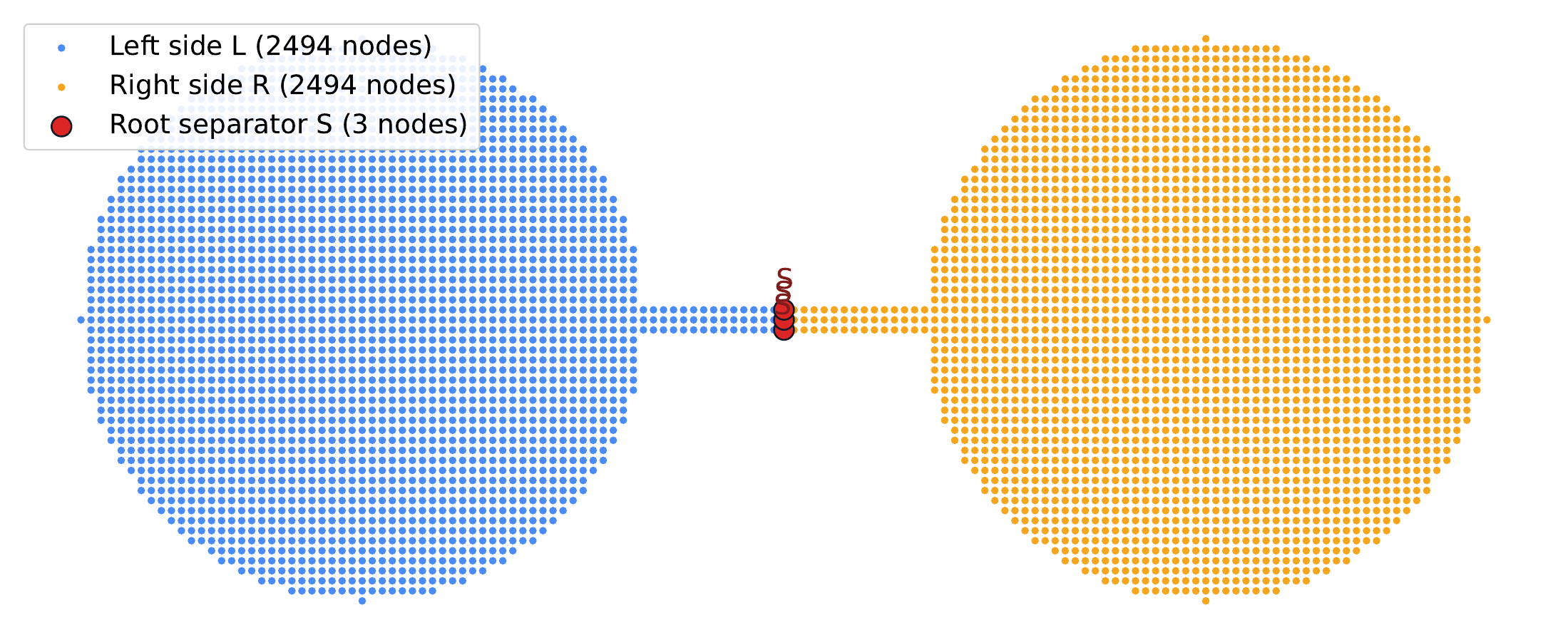}
    \caption{\small{Planar dumbbell graphs after the root separator split. The two lobes form the left and right subgraphs, while the narrow bridge induces a small separator; the examples above correspond to two different handle widths.}}
    \label{fig:experiments_dumbbell_graph}
\end{figure}

To build some intuition for how GenusSink constructs the separation tree and applies the kernel, we begin by considering a particularly stylized example, that is, a dumbbell-shaped planar graph, shown in Fig.~\ref{fig:experiments_dumbbell_graph}. This geometry admits a small separator that splits the graph into two nearly equal subgraphs, making it a natural setting for a depth-1 separation tree.

\begin{figure}[h]
    \centering
    \includegraphics[width=0.9\linewidth]{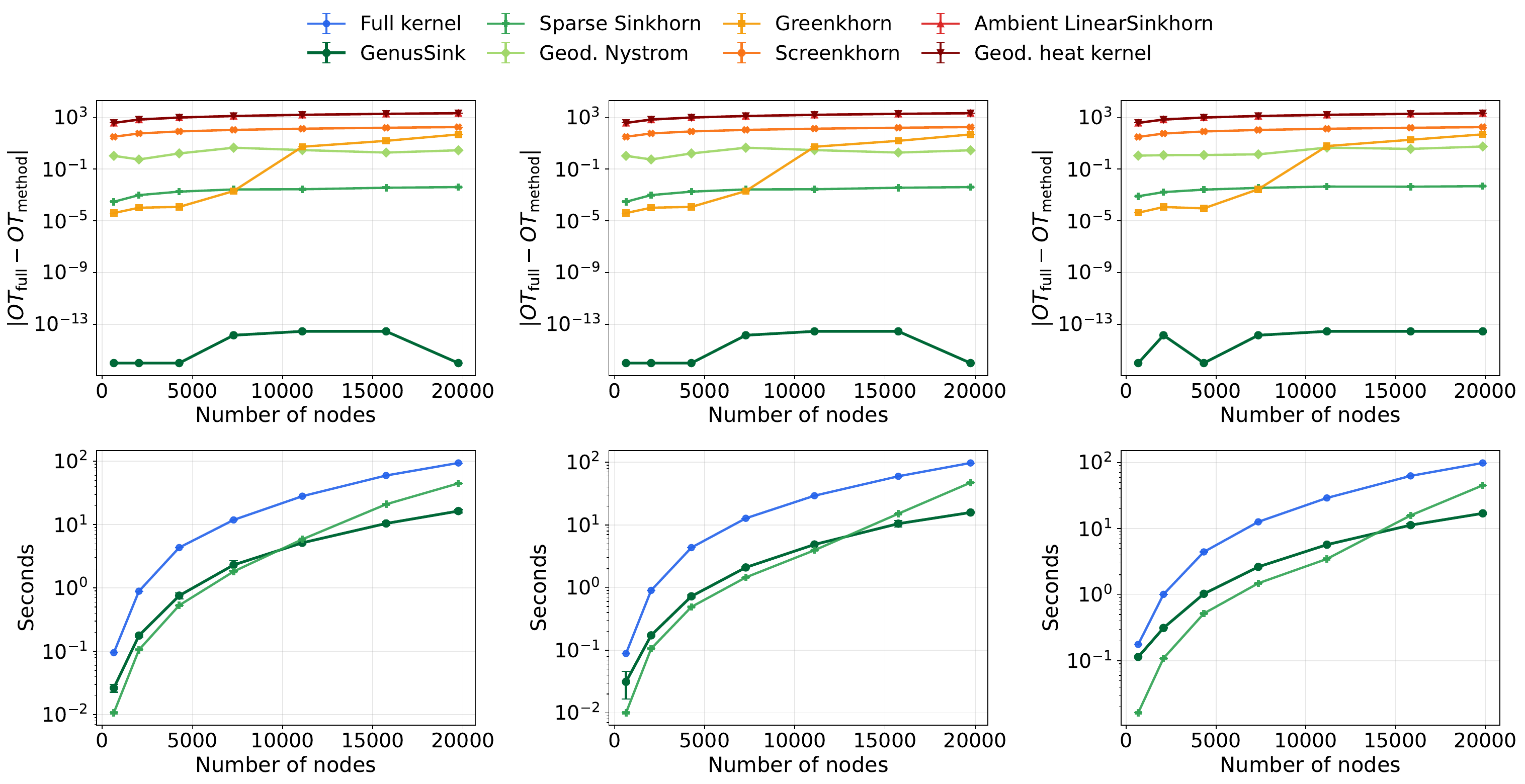}
    \caption{\small{Planar dumbbell experiments for handle widths $w=1,2,3$ (left to right). \textbf{Top:} absolute Sinkhorn-cost error relative to the dense geodesic full-kernel baseline. \textbf{Bottom:} total runtime for the baseline and two most accurate efficient methods: Sparse Sinkhorn and GenusSink. Across all widths, GenusSink remains numerically exact while providing significant computational gains.}}
    \label{fig:experiments_dumbbell_graph_results}
\end{figure}

We evaluate the time required for Sinkhorn to converge to stopping tolerance $\tau = 10^{-7}$ on dumbbell graphs with radii $r \in \{10,18,26,34,42,50,56\}$, bridge length equal to the radius, handle width $w \in \{1,2,3\}$, and initial/target distributions given by two-component Gaussian mixtures localized on the left and right lobes, respectively, and normalized over the graph vertices. The entropic regularization parameter is set to $\varepsilon = 0.3r$. Since the handle width also controls the separator size in this construction, these experiments illustrate how the performance of GenusSink changes as the separator becomes larger. We compare against the dense full-kernel geodesic Sinkhorn baseline, Greenkhorn~\cite{sinkhorn-2}, Screenkhorn~\cite{alaya2019screenkhorn}, Sparse Sinkhorn~\cite{schmitzer2019stabilized}, Geodesic heat kernel~\cite{huguet2023geodesic}, and two Linear Sinkhorn baselines~\cite{linear-time-sinkhorn-positive-features}: the original ambient-Euclidean feature construction and a geodesic Nystr\"om approximation. With separation height fixed to $1$, GenusSink converges substantially faster than the brute-force baseline that materializes the full dense kernel and then performs matrix--vector products; see Fig.~\ref{fig:experiments_dumbbell_graph_results}. As the graph size grows, constructing the dense kernel becomes increasingly expensive due both to memory costs and to the need to compute all pairwise shortest-path distances. In contrast, GenusSink remains exact up to floating-point precision, whereas the approximate baselines incur visibly larger error.

\begin{figure}[h]
    \centering
    \includegraphics[trim={0cm 4cm 0cm 20cm}, clip, width=1\linewidth]{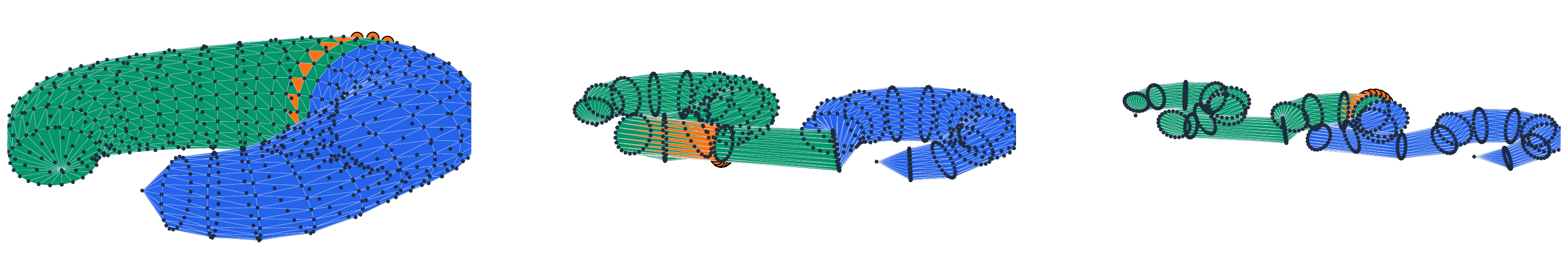}
    \caption{\small{Pseudo-genus surface family used in our synthetic experiments. We show planar sphere-topology surface meshes embedded in $\mathbb{R}^3$ for pseudo-genus orders 1, 2, and 3, colored according to the root separator split used to initialize the separation tree.}}
    \label{fig:pseudo-genus-meshes-3d}
\end{figure}

In addition to the dumbbell graph, we evaluate our method on a family of boundaryless sphere-topology ``pseudo-genus'' surface graphs embedded in $\mathbb{R}^3$ (as seen in Section~\ref{sec:experiments}). These graphs remain planar, but become progressively more complicated geometrically. Specifically, increasing the pseudo-genus order introduces additional narrow, highly squashed neck regions that visually resemble higher-genus shapes while preserving planarity, which is essential for our separator-tree construction. The resulting family therefore provides a more challenging synthetic testbed than the dumbbell graph while still lying in the regime targeted by our method (see Fig.~\ref{fig:pseudo-genus-meshes-3d}).

For these timing and accuracy experiments, we use pseudo-genus graphs of orders $1$, $2$, and $3$, with node counts matched to the width-$1$ dumbbell size sequence, namely $n \in \{644, 2036, 4268, 7283, 11090, 15740, 19745\}$. On each graph, we compute Sinkhorn transport between two smooth nonuniform probability distributions, defined as two-component Gaussian mixtures in the surface parameter coordinates and normalized over the graph vertices. We again compare GenusSink against the dense geodesic full-kernel Sinkhorn baseline, Greenkhorn, Ambient LinearSinkhorn, Geod. Nystr\"om, Sparse Sinkhorn, Screenkhorn, and Geod. heat kernel. The approximate baselines avoid the exact separator-tree computation but incur visible transport-cost error; in particular, Ambient LinearSinkhorn uses Euclidean features rather than the intrinsic graph geometry. 

\subsection{Details using Thingi10K data}\label{app:details-thingi10k}

As mentioned in Section \ref{sec:experiments}, we evaluate GenusSink on a variety of Thingi10K triangle meshes converted to weighted graph instances. For each mesh, every unique triangle edge is used as an undirected graph edge with weight equal to its Euclidean length. The experiment uses 28 meshes, with mesh IDs
\[
\begin{gathered}
636808,472186,52562,636795,636813,149255,636794,650181,51809,642501,\\
650180,39879,131453,650183,636818,650184,79741,780015,636800,636796,\\
636816,650185,636799,636817,636797,636803,159695,41272.
\end{gathered}
\]
The corresponding vertex counts range from \(2771\) to \(12472\).  Source and target measures are deterministic smooth geodesic Gaussian mixtures. PCA anchors define left, right, low, and high mesh locations; the source uses anchors \((\mathrm{left},\mathrm{low})\) with weights \((0.7,0.3)\), while the target uses \((\mathrm{right},\mathrm{high})\) with weights \((0.65,0.35)\). For graph distance \(d_G\), each mixture component is proportional to
\[
\exp\!\left(-\frac{d_G(x,c)^2}{2\sigma^2}\right),
\qquad
\sigma=\max\{0.18 \times\mathrm{diam},\;3\,\overline{w},\;10^{-8}\},
\]
where \(\overline{w}\) is the mean mesh-edge length and $\mathrm{diam}$ is the Euclidean diagonal length of the mesh's axis-aligned bounding box. The entropic scale is \(\varepsilon=0.2\times\mathrm{diam}\).  
We compare against dense geodesic Sinkhorn as the reference and report both dense kernel construction time versus GenusSink total runtime, and absolute OT-cost error relative to the dense reference.  Additional accuracy baselines are Greenkhorn with \(50{,}000\) updates and tolerance \(10^{-4}\), Geodesic Nyström with rank \(64\), Ambient LinearSinkhorn with \(64\) random features, Screenkhorn with budget fraction \(0.5\) and \(\texttt{maxiter}=\texttt{maxfun}=500\), sparse truncated Sinkhorn retaining entries with \(K_{ij}\ge 10^{-4}\), and geodesic heat-kernel Sinkhorn with \(\tau=0.25\varepsilon\), Chebyshev order \(20\), and inverse-distance graph Laplacian weights.

\subsection{Broader Impact \& Limitations}
\label{sec:lim}
\paragraph{Broader Impact.} This paper might potentially have a positive impact on the new classes of Transformer architectures that leverage Optimal Transport methods in their attention mechanisms, by making these mechanisms much more computationally efficient.

\paragraph{Societal Impact.} We are not aware of any societal impact that this paper might have, as it is mainly of the algorithmic flavor, proposing a new class of efficient Sinkhorn algorithms operating on bounded genus graphs and with cost functions defined via shortest-path-distance metrics.
\paragraph{Limitations.} This paper focuses on bounded genus graphs. In future work, we plan to extend to larger classes of graphs, in particular for minor-free graphs (every bounded genus graph is minor-free, and the class of minor-free graphs is a strict super-class of the bounded genus graphs).

\end{document}